\documentclass{article}

\usepackage{arxiv}

\usepackage[utf8]{inputenc} 
\usepackage[T1]{fontenc}    
\usepackage{hyperref}       
\usepackage{url}            
\usepackage{booktabs}       
\usepackage{amsfonts}       
\usepackage{amsmath}        
\usepackage{nicefrac}       
\usepackage{microtype}      
\usepackage{lipsum}
\usepackage{graphicx}
\usepackage{bm}
\usepackage[square,numbers,sort&compress]{natbib}

\usepackage[usenames,dvipsnames]{xcolor}
\newcommand{\hlc}[2][red]{{\textcolor{black}{#2}}}

\makeatletter
\def\def@donotrepeattitle{}
\makeatother

\title{\Large Data-driven discovery of dynamics from time-resolved coherent scattering}
\date{}

\usepackage{authblk}
\author[1*]{Nina Andrejevic}
\author[1]{Tao Zhou}
\author[2]{Qingteng Zhang}
\author[2]{Suresh Narayanan}
\author[2]{Mathew J. Cherukara}
\author[1]{Maria K. Y. Chan}
\affil[1]{\textit{Center for Nanoscale Materials, Argonne National Laboratory, Lemont, IL 60439, USA}}
\affil[2]{\textit{Advanced Photon Source, Argonne National Laboratory, Lemont, IL 60439, USA}}
\affil[*]{E-mail: \texttt{nandrejevic@anl.gov}}
\begin{document}

\maketitle
\begin{abstract}
Coherent X-ray scattering (CXS) techniques are capable of interrogating dynamics of nano- to mesoscale materials systems at time scales spanning several orders of magnitude. However, obtaining accurate theoretical descriptions of complex dynamics is often limited by one or more factors -- the ability to visualize dynamics in real space, computational cost of high-fidelity simulations, and effectiveness of approximate or phenomenological models. In this work, we develop a data-driven framework to uncover mechanistic models of dynamics directly from time-resolved CXS measurements without solving the phase reconstruction problem for the entire time series of diffraction patterns. Our approach uses neural differential equations to parameterize unknown real-space dynamics and implements a computational scattering forward model to relate real-space predictions to reciprocal-space observations. This method is shown to recover the dynamics of several computational model systems under various simulated conditions of measurement resolution and noise. Moreover, the trained model enables estimation of long-term dynamics well beyond the maximum observation time, which can be used to inform and refine experimental parameters in practice. Finally, we demonstrate an experimental proof-of-concept by applying our framework to recover the probe trajectory from a ptychographic scan. Our proposed framework bridges the wide existing gap between approximate models and complex data.
\end{abstract}


\section{Introduction}
Coherent X-ray scattering (CXS) techniques play a critical role in the investigation of nano- and mesoscale materials systems significant to a wide variety of scientific disciplines and applications. Owing to continued advancements in source brightness, instrumentation, and detector capabilities at synchrotron X-ray facilities, alongside complementary iterative algorithms, CXS-based imaging techniques such as coherent diffraction imaging (CDI), including Bragg CDI (BCDI) and ptychography \cite{yao2021method}, have enabled the study of materials at unprecedented levels of detail, such as imaging of nanoscale strain and defect displacement fields \cite{robinson2009coherent,hruszkewycz2012quantitative,clark2015three,ulvestad2015topological,ulvestad2016situ,choi2020situ}, ferromagnetic and ferroelectric domain patterns \cite{tripathi2011dichroic,hruszkewycz2013imaging,hruszkewycz2016structural}, and extended objects like integrated circuits \cite{jiang2021achieving,holler2019three}, among many others. However, these methods depend on the ability to invert diffraction patterns into real-space images using iterative phase retrieval algorithms \hlc{by exploiting} certain support or overlap constraints. This typically limits dynamic imaging with coherent X-rays to isolated objects that fit within the illuminating beam \cite{clark2013ultrafast,hinsley2020dynamic} or to materials systems with comparatively slow dynamics. \hlc{Approaches to accelerate dynamic coherent imaging have been proposed on several fronts. For instance,} recent innovations in beam optics have indicated the potential for single-shot CDI and ptychography with nearly 50 nm spatial resolution using 10 ms exposures \cite{takazawa2021demonstration,takazawa2023coupling}, and $\sim$5 $\mu$m spatial resolution using femtosecond pulses from a soft X-ray free-electron laser \cite{kharitonov2022single}. \hlc{Additional spatiotemporal constraints have also been introduced either by identifying static regions of the sample computationally \cite{hinsley2020dynamic} or by employing a dual-pinhole aperture to concurrently probe dynamic and static regions of the sample plane \cite{lo2018situ}. However, inverting an entire time series of diffraction patterns frame-by-frame can be costly and is fundamentally distinct from the process of modeling the underlying dynamics, which is necessary to forecast evolving materials behavior.}

Conversely, X-ray photon correlation spectroscopy (XPCS) \cite{grubel2008x,shpyrko2014x} is a CXS technique that exploits correlations between time-dependent fluctuations of the scattered intensity to derive insights about sample dynamics and is accessible at time scales spanning several orders of magnitude \cite{lehmkuhler2021femtoseconds}. The versatility of this approach has enabled the study of dynamic phenomena across a broad range of materials systems \cite{zhang2018dynamics,sandy2018hard}, including colloidal suspensions \cite{dierker1995x,fluerasu2010dynamics,pal2018anomalous}, glasses \cite{ruta2012atomic,evenson2015x,kwasniewski2014anomalous}, and (anti)ferromagnetic and ferroelectric domains and textures \cite{shpyrko2007direct,konings2011magnetic,chen2013jamming,lim2014coherent,seaberg2017nanosecond,zhang2017thermal,gorfman2018ferroelectric,ricci2020intermittent}. The analysis of intensity-intensity correlation functions yields characteristic time scales of the underlying dynamical process which can inform the development of mechanistic models; however, insights from these measurements are therefore more indirect compared to imaging methods which visualize dynamics in real space. As a result, theory development can be nontrivial for complex dynamics which differ significantly from simple diffusive behavior, particularly in out-of-equilibrium conditions \cite{madsen2010beyond}. This raises a need for methodologies capable of bridging the wide existing gap between approximate theoretical models and the often complicated dynamic events behind complex data.

Rather than interpret measurements through the lens of approximate models, this work aims to derive mechanistic insights about dynamics directly from data through a combination of machine learning and scientific computing techniques. The intersection of modern machine learning with scientific computing is often termed scientific machine learning, encompassing methods such as physics-informed neural networks \cite{karniadakis2021physics,raissi2018hidden,raissi2018deep,raissi2019physics} and neural differential equations \cite{chen2018neural,rackauckas2020universal,chen2020learning,shankar2020learning,zubov2021neuralpde}. These techniques leverage the flexibility of machine learning models while ensuring their predictions adhere to physical principles or other domain knowledge; as a result, they have shown to be more interpretable and robust in a wide range of scientific contexts \cite{rao2021embedding,lu2022discovering,richter2022neural,linot2023stabilized,wang2023learning,covington2022bridging}. Moreover, advancements in sparsity-exploiting methods such as symbolic regression and sparse identification of nonlinear dynamics (SINDy) \cite{brunton2016discovering} suggest the possibility to derive governing equations in the form of simple mathematical expressions from highly complex observations \cite{rudy2017data,champion2019data}.

In this work, we develop a data-driven framework that parameterizes unknown real-space dynamics using neural differential equations. These networks are optimized through calculation of coherent diffraction patterns from the predictions of real-space coordinates at each iteration, which are compared against the observations from time-resolved coherent scattering, either simulated or experimental. Importantly, this approach enables the discovery of dynamics without inversion of the entire time series of diffraction patterns. Instead, only real-space information at the initial conditions is needed in practice, which can be obtained, for example, through traditional CDI or ptychography under static conditions. This opens up the possibility of visualizing inaccessible, real-space dynamics evolving faster than typical CDI or ptychography experiments by relying only on a few high-fidelity measurements to establish the initial conditions and inferring subsequent dynamics from fast scans that are too challenging for phase retrieval. Additionally, the trained neural networks \hlc{directly model the dynamics themselves and can therefore be used} to extrapolate well beyond the training window, which can critically inform experimental design and planning. The main outcome of this work is thus a general and versatile platform for data-driven discovery of dynamics in the context of time-resolved CXS.

\section{Results}
\begin{figure}[t]
    \centering
    \includegraphics[width=\textwidth]{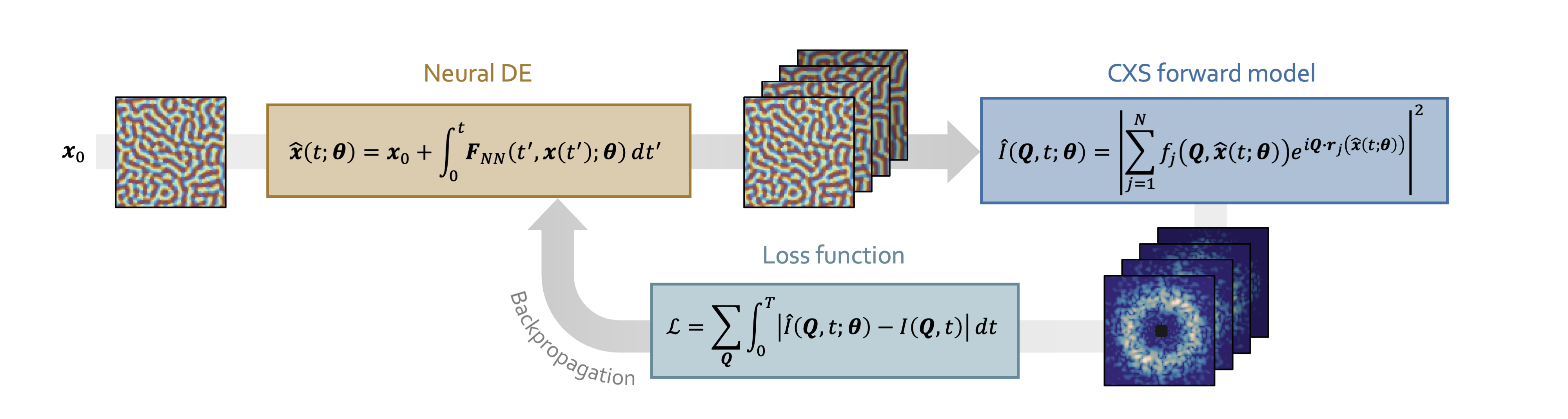}
    \caption{\textbf{Data-driven modeling of dynamics.} Given the initial system state $\bm{x}_0$, the neural differential equation (DE) $\bm{F}_{\text{NN}}(t,\bm{x}(t);\bm{\theta})$, parameterized by trainable weights $\bm{\theta}$, is integrated numerically to obtain the predicted time evolution of the system state $\bm{\hat{x}}(t;\bm{\theta})$. The corresponding scattered intensity $\hat{I}(\bm{Q},t;\bm{\theta})$ is obtained by applying the CXS forward model to the predicted $\bm{\hat{x}}(t;\bm{\theta})$. The loss function compares the predicted intensity against the target intensity $I(\bm{Q},t)$, and the network weights $\bm{\theta}$ are updated through backpropagation.}
    \label{fig:fig1}
\end{figure}

Figure \ref{fig:fig1} illustrates the general workflow for the data-driven dynamical modeling approach implemented in this work. For a given system, the unknown real-space dynamics are parameterized by neural differential equations, for instance,
\begin{equation}
    \frac{d\bm{x}(t)}{dt} \approx \bm{F}_{\text{NN}}(t,\bm{x}(t);\bm{\theta}),
\end{equation}
where $\bm{x}(t)$ are the dynamic coordinates of interest, and $\bm{\theta}$ represents the set of trainable neural network weights. We use the term dynamic coordinates to refer to real-space variables that describe the state of the system at a given time and whose evolution is modeled by unknown differential equations. For instance, for a system of dynamic particles with known geometries, the dynamic coordinates can simply refer to particle positions; however, the dynamic coordinates can refer more generally to collective variables and even individual image pixels visualizing a physical quantity such as local electron density. Given the initial system state $\bm{x}_0$ specified by the chosen dynamic coordinates, $\bm{F}_{\text{NN}}$ can be numerically integrated in time to obtain the predicted time-evolution of the system, $\bm{\hat{x}}(t;\bm{\theta})$. Normally, the predicted result is compared directly with the expected $\bm{x}(t)$ at discrete time points using a chosen loss function \cite{chen2018neural}. However, as the real-space dynamics are not directly accessible, we further incorporate a computational scattering forward model to relate the real-space predictions to reciprocal-space observations -- the scattered intensities $I(\bm{Q},t)$. In particular, we compute the elastic scattered intensity $\hat{I}(\bm{Q},t;\bm{\theta})$ from a coherent X-ray probe incident on the scattering medium with predicted coordinates $\bm{\hat{x}}(t;\bm{\theta})$. Details of the coherent scattering calculation are provided in the Methods. Finally, the loss $\mathcal{L}$ is defined in terms of the predicted and observed scattered intensities in reciprocal space, $\hat{I}(\bm{Q},t;\bm{\theta})$ and $I(\bm{Q},t)$, respectively.

In the following, we present several computational case studies to evaluate the performance of this framework in the context of different physical systems and experimental scenarios with increasing complexity. \hlc{In each case study, we evaluate the performance of the framework using an ensemble of 10 independent neural network models trained using different sets of initial parameter weights.} We then establish a proof-of-concept experiment to assess the adaptation of the framework to an experimental setting.

\subsection{Computational case studies}
\begin{figure}[t]
    \centering
    \includegraphics[width=\textwidth]{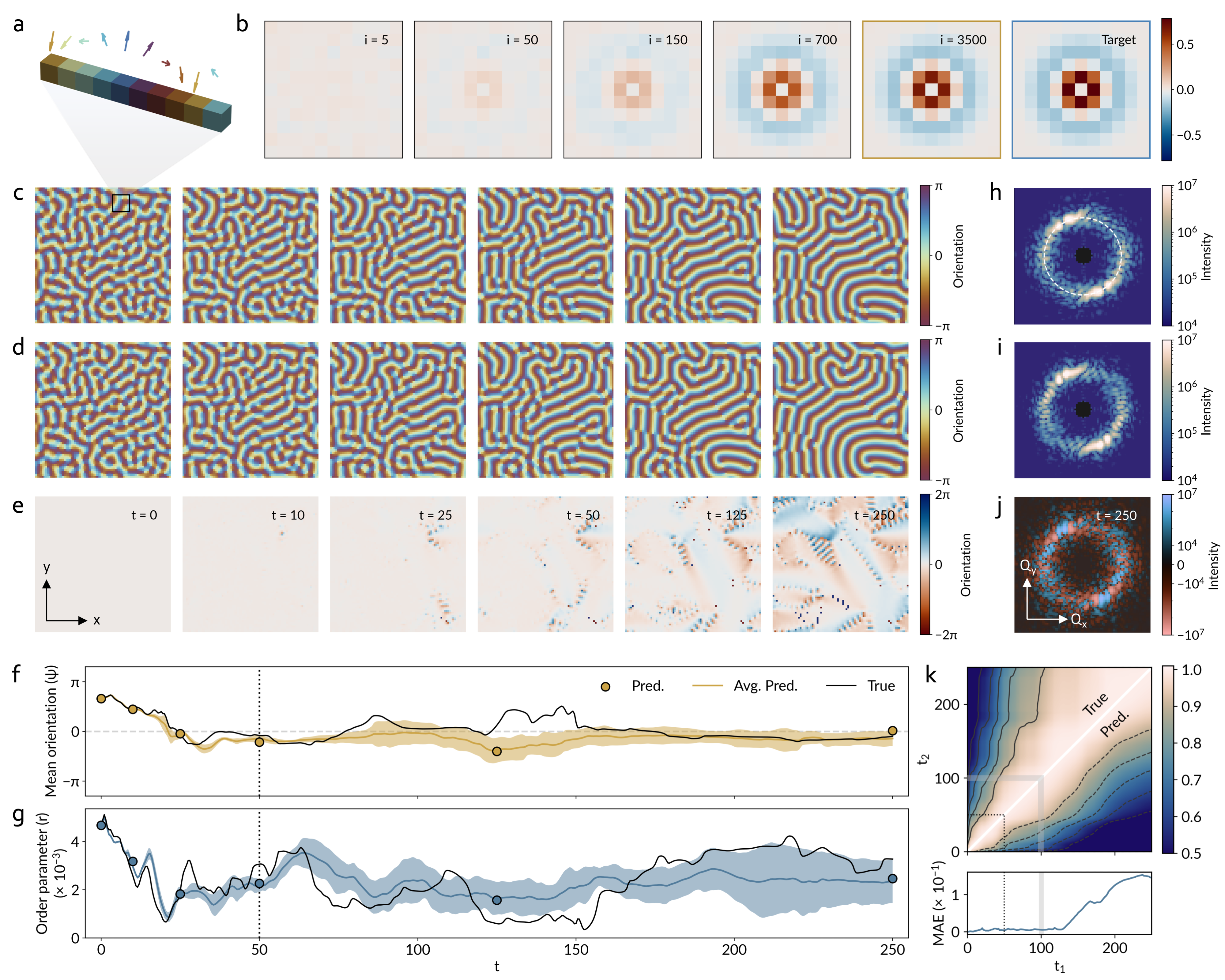}
    \caption{\textbf{Case study 1: Synchronization dynamics of locally-coupled moments.} \textbf{a.} Schematic representation of locally-coupled moments colored by orientation, collectively exhibiting a stripe domain pattern. \textbf{b.} Evolution of the learned spatial coupling kernel \hlc{of a representative model} during training, depicted after 5, 50, 150, 700, and 3500 iterations. Panels outlined in yellow and blue represent the final predicted and ground truth kernels, respectively. \textbf{c-e.} From top to bottom, the time evolution of the true state, predicted state, and difference map of the system at selected time points \hlc{for one representative model and initial condition}. The final time point ($t=250$) corresponds to five times the maximum time seen during training. \hlc{\textbf{f-g.} The time evolution of the mean orientation, $\psi$ (\textbf{f}), and order parameter, r (\textbf{g}), of the system in \textbf{c} and \textbf{d}. Solid colored lines correspond to the model ensemble average, and shading represents one standard deviation. Scattered points correspond to the 6 panels in \textbf{d}.} \textbf{h-j.} From top to bottom, the true (\textbf{h}) and predicted (\textbf{i}) diffraction patterns and their difference map (\textbf{j}) corresponding to the system state at $t=250$. The black central pixels correspond to the masked, unscattered beam and are not factored into loss calculations. \hlc{\textbf{k.} The two-time intensity-intensity correlation functions calculated from the diffraction patterns of the true (upper triangle) and average predicted (lower triangle) systems. The correlation functions were calculated by taking an azimuthal average over each diffraction pattern at the equivalent $Q$ indicated by the white dashed line in \textbf{h}. The lower panel shows the MAE between slices of the true and predicted two-time maps as a function of time; and example slice is indicated by the gray shaded rectangles. Black dotted lines denote the maximum time seen during training.}}
    \label{fig:fig2}
\end{figure}

\paragraph{Case study 1: Locally-coupled moments}
We consider a system of $N$ moments with fixed positions $\bm{r}_j$ on a square two-dimensional lattice (Figure \ref{fig:fig2}\textbf{a}), each characterized by a form factor $f(\varphi_j(t))$ that is dependent upon the dynamic moment orientation $\varphi_j(t)$. In our simulations, the time evolution of moments is modeled by the Kuramoto model \cite{kuramoto1984springer}, which couples the time evolution of neighboring moments according to,
\begin{equation}
    \frac{d\varphi_i(t)}{dt} = \sum_{j \in \mathcal{N}(i)}a_{ij}\sin{(\varphi_j(t) - \varphi_i(t))},
\end{equation}
where $\mathcal{N}(i)$ denotes the neighborhood of the $i^{\text{th}}$ moment and $a_{ij}$ denotes the pairwise coupling strength. We begin by approximating this ordinary differential equation (ODE) with a convolutional neural network (CNN) in which the trainable weights \hlc{of the convolution kernel} parameterize the coupling kernel $\bm{a}$. However, rather than directly observe the evolution of moments in real space, the network is trained by comparing predicted and calculated diffraction patterns through a chosen loss function, here selected as the mean absolute error (MAE). The diffraction patterns were passed to the network in mini-batches sampled from 100 time series with different initial configurations. Each \hlc{of these time series} consisted of an $80 \times 80$ grid of moments and an $11 \times 11$ coupling kernel, simulated up to a maximum time $t=50$ at time intervals of $0.5$. \hlc{Each random mini-batch consisted of 20 different simulations spanning a time interval of 15 (\textit{i.e.}, 30 time steps), thereby sampling initial conditions at different points in the dynamics up to $t=35$.} Additional details of the coherent scattering calculation are provided in the Methods. Figure \ref{fig:fig2}\textbf{b} shows the evolution of \hlc{a representative} coupling kernel \hlc{learned} over the course of training. We can see evidence of the expected symmetry in the coupling kernel emerge after approximately 150 iterations and continuously approach the target thereafter. \hlc{The training histories of the coupling kernels in the 10-model ensemble are plotted in \hlc{Supplementary} Figure \ref{fig:figS2}.} A representative example \hlc{among 50 different initial configurations is selected to show} the time evolution of true and predicted system states in real space in Figures \ref{fig:fig2}\textbf{c} and \textbf{d}, respectively. We find good agreement between the true and predicted states and diffraction patterns up to $t=250$ (Figures \ref{fig:fig2}\textbf{h} and \textbf{i}), five times the maximum time seen during training, indicating that the neural ODE extrapolates well beyond the training window.

The difference maps between the true and predicted real- and reciprocal-space images are shown in Figures \ref{fig:fig2}\textbf{e} and \textbf{j}. \hlc{We note that errors in the real-space images tend to originate and propagate along specific paths of the domain which appear to coincide with points of large vorticity, defined here as $\nabla \times \left< \cos{(\varphi_j)}, \sin{(\varphi_j)} \right >$; this is examined in greater detail in \hlc{Supplementary} Figures \ref{fig:figS3}\textbf{a} and \textbf{b}. Such a finding appears intuitive, as these points have the highest local variation in the moment orientations. The predictions also tend to evenly over- and underestimate the moment orientations in a given frame, evidenced by the error distributions of the real-space difference maps plotted in \hlc{Supplementary} Figure \ref{fig:figS3}\textbf{d}. This motivates the comparison of orientation statistics between true and predicted systems as a function of time. Following Ref. \citenum{breakspear2010generative}, we compute the centroid vectors of the moment distributions,
\begin{equation}
    \hlc{r(t)e^{i\psi(t)} = \frac{1}{N}\sum_{j=1}^N e^{i\varphi_j(t)},}
\end{equation}
where $\psi(t)$ is the average moment orientation and $r(t)$ is an order parameter quantifying the degree of uniformity in the moments. Figures \ref{fig:fig2}\textbf{f} and \textbf{d} show the calculated $\psi$ and $r$ for the example visualized in Figures \ref{fig:fig2}\textbf{c} and \textbf{d}, indicating relatively good agreement. The discrepancy in $\psi$ between $t=125$ and $t=150$ is potentially caused by the fact that the coherent scattering calculation is an even function of the moment orientations, leading to a potential ambiguity about 0. Alternately, as this discrepancy coincides with a reduction in the true order parameter, $r$, it may be due to local changes in the ground truth dynamics that occur only later in the prediction. However, it is worth noting that the order parameter, which has possible values between 0 and 1, is consistently small overall as the moments are never fully synchronous due to the choice of coupling kernel, which leads to the characteristic formation of striped domains. As stated, the example shown in Figure \ref{fig:fig2} is representative of the average performance of the model ensemble across 50 different initial configurations (\hlc{Supplementary} Figure \ref{fig:figS4}\textbf{a}). Simulations which were comparatively less and more challenging to predict accurately, as given by the real-space MAE, are shown in \hlc{Supplementary} Figures \ref{fig:figS5} and \ref{fig:figS6}, respectively.}

\hlc{Regarding the evaluation of errors in reciprocal space, it is worth comparing the diffraction patterns obtained from the true and predicted dynamics in terms of not only the loss function used for training, but also the expected intensity-intensity correlations that would be obtained in an XPCS setting. Specifically, we compute the two-time correlation function, $C_2(Q, t_1, t_2)$ \cite{bikondoa2017use,brown1997speckle}, of the true and predicted diffraction patterns for each of the 10 models in the ensemble. These correlation functions were calculated by taking the azimuthal averages at the equivalent $Q$ indicated by the white dashed line in Figure \ref{fig:fig2}\textbf{h}. Additional details of the $C_2$ calculations are provided in the Methods. Figure \ref{fig:fig2}\textbf{k} compares the true $C_2$ against the average $C_2$ obtained from the 10-model ensemble. We find very good agreement until approximately $t=125$, at which point the predicted dynamics remain slightly faster than the true, though both continue to progressively slow down as the moment orientations synchronize. This is corroborated by the values of the correlation time $\tau$ as a function of $t_1$, which we obtain by fitting the one-time correlation functions extracted at fixed observation times (\hlc{Supplementary} Figure \ref{fig:figS4}\textbf{d}) \cite{bikondoa2017use}. Since the networks are trained on observations at early times when dynamics are generally faster, they may be more likely to slightly overestimate the rates of dynamics in a slower regime, particularly when the onset is abrupt. The divergence in $C_2$ also approximately coincides with the discrepancy in mean orientations discussed earlier, which is supported by the fact that the true dynamics begin to slow down more rapidly than the predicted around this time. Though beyond the scope of the present work, directly incorporating $C_2$ into the loss function may be able to further strengthen these results.}

\hlc{Finally, we note that the CNN model can successfully learn the local coupling kernel even when the trainable weights overparameterize the true kernel, \textit{i.e.}, we allow for a receptive field greater than $11 \times 11$ in the present example. As an example, \hlc{Supplementary} Figure \ref{fig:figS7} shows the results of training on the same dataset but with a convolution kernel of size $21 \times 21$, which also converges to the expected result.}

\begin{figure}[t]
    \centering
    \includegraphics[width=\textwidth]{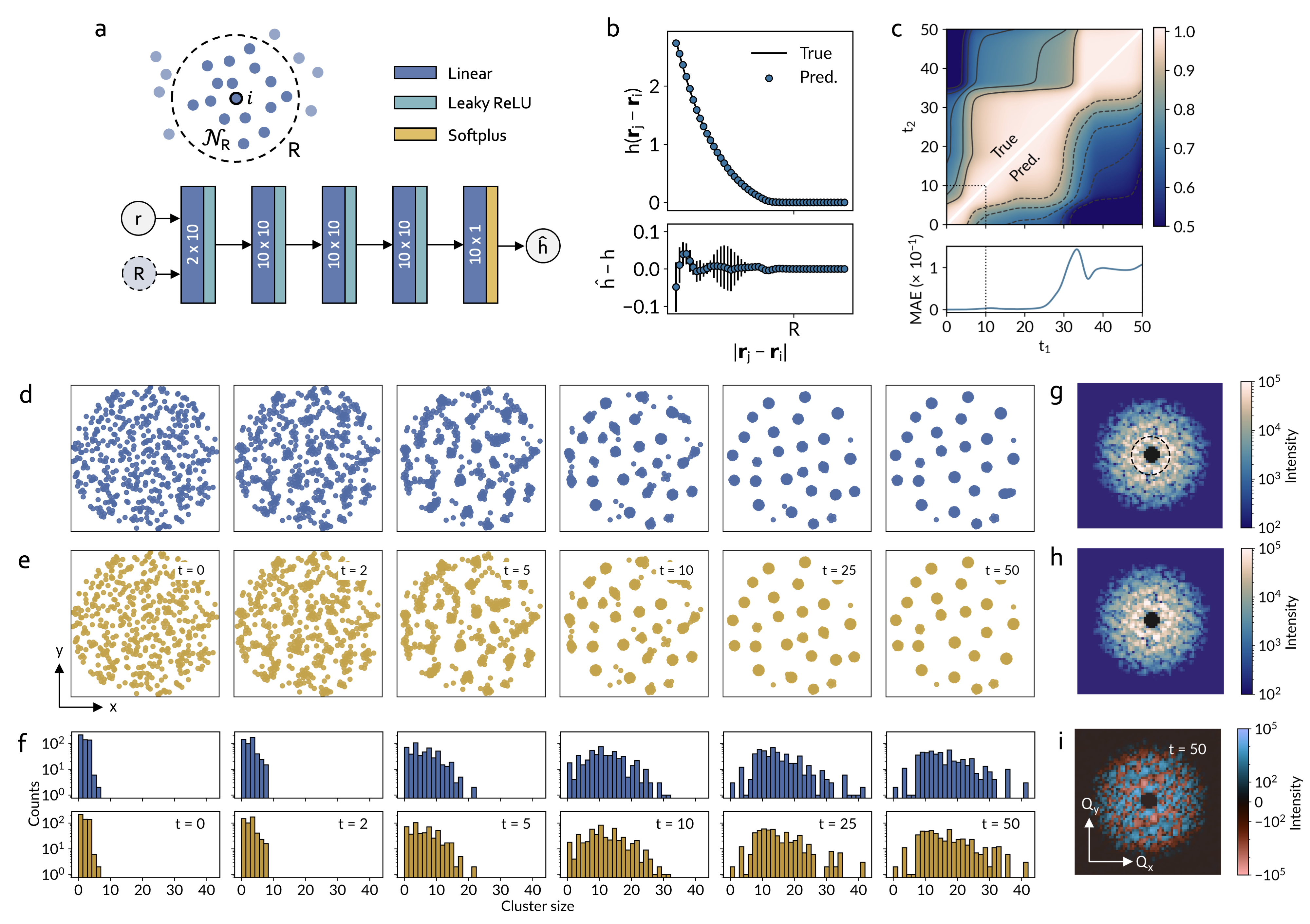}
    \caption{\textbf{Case study 2: Self-organizing behavior in a system of interacting particles.} \textbf{a.} Schematic of the neural network architecture used to model the smooth cutoff function governing particle interactions. \hlc{Top left illustration shows the definitions of the number of neighboring particles $\mathcal{N}_R$ within a radius $R$ of the $i^{th}$ particle.} \textbf{b.} The true (solid line) and \hlc{ensemble-average} predicted (filled circles) cutoff functions. \hlc{The bottom panel plots the difference between the two in order to better visualize the error bars, which represent one standard deviation among the 10-model ensemble.} \textbf{c.} The two-time intensity-intensity correlation function calculated from the diffraction patterns of the true \hlc{(upper triangle)} and predicted \hlc{(lower triangle)} systems. The correlation functions were calculated by taking an azimuthal average over each diffraction pattern at the equivalent $Q$ indicated by the \hlc{black} dashed line in \textbf{g}. \textbf{d-e.} The time evolution of the true (\textbf{d}) and predicted (\textbf{e}) states of the system at selected time points \hlc{for one representative model and initial condition}. The final time point ($t = 50$) corresponds to five times the maximum time seen during training. \textbf{f.} The corresponding time evolution of the true (top) and predicted (bottom) cluster size distributions. \textbf{g-i.} From top to bottom, the true (\textbf{g}) and predicted (\textbf{h}) diffraction patterns and their difference map (\textbf{i}) corresponding to the system state at $t = 50$.}
    \label{fig:fig3}
\end{figure}

\paragraph{Case study 2: Self-organizing particles}
To evaluate the performance of the framework on more complex learning tasks beyond parameter estimation, we next consider clustering dynamics in a collection of interacting particles with full spatial degrees of freedom in which the unknown dynamics are governed by a continuous-valued function. Following the work of Ref. \citenum{o2017oscillators}, we model the evolution of each particle's spatial coordinates, $\mathbf{r}_i$, according to,
\begin{equation}
    \frac{d\mathbf{r}_i(t)}{dt} = \frac{1}{\hlc{n}}\sum_{j \in \mathcal{N}_R(i)} h(\mathbf{r}_j - \mathbf{r}_i)\left(1 - h(\mathbf{r}_j - \mathbf{r}_i) \right) \left( \mathbf{r}_j - \mathbf{r}_i \right),
\end{equation}
where $\mathcal{N}_R$ enumerates particles found within a cutoff radius $R$ of the $i^{th}$ particle, $h(\mathbf{r}_j - \mathbf{r}_i)$ is a smooth cutoff function which determines the interaction potential, and \hlc{$n$} is a normalizing factor \hlc{which scales} with the size of the simulation box, $L$, \hlc{as $n=(2/L)^2$}. To obtain a system which exhibits clustering, we model the cutoff function as a short-range interaction according to \cite{agueh2011analysis},
\begin{equation}
    h(r) =  
    \begin{cases}
      \frac{R}{2r}\left(1 + \tanh{\left(\frac{1}{r - R} - \frac{1}{r - 2R} \right)} \right) & r \leq R\\
      0 & r > R,
    \end{cases}
\end{equation}
where $r = |\mathbf{r}|$. In this case study, we simulate \hlc{dynamics from 50 different initial configurations, each comprising} $500$ identical particles with a radius of $0.03L$ and with a cutoff radius of $R=0.15L$ for particle interactions. We parameterize the cutoff function using the neural network illustrated in Figure \ref{fig:fig3}\textbf{a}, which is trained through observation of the calculated diffraction patterns from \hlc{mini-batches of 4 different initial configurations at a time. Each mini-batch is sampled from the full simulation time of} $t=10$ \hlc{in batch times of 2} at time intervals of $0.1$ \hlc{(\textit{i.e.}, 20 time steps per batch)}. Because the particle interactions are governed by a shared pairwise potential, we implement this network in the form of a graph neural ODE,
\begin{equation}
    \frac{d\mathbf{r}_i(t)}{dt} \approx \frac{1}{\hlc{n}}\sum_{j \in \mathcal{N}_R(i)} F_{\text{NN}}(t,r; R, \bm{\theta})\left(1 - F_{\text{NN}}(t,r; R, \bm{\theta}) \right)\left(\mathbf{r}_j - \mathbf{r}_i \right),
\end{equation}
where $F_{\text{NN}}$ denotes the neural network, $R$ is a trainable parameter representing the cutoff radius, and $\bm{\theta}$ represents the remaining trainable weights of the neural network. Here, $F_{\text{NN}}$ is a feed-forward neural network comprising the input and output layers and three hidden layers each with 10 neurons (Figure \ref{fig:fig3}\textbf{a}). Each layer prior to the output is activated by a leaky rectified linear unit (ReLU), and a softplus is used as the final activation to smoothly enforce positivity. At each iteration of the training loop, a set of graphs is constructed based on the neighborhood defined by the current estimate of the cutoff radius, $R^{(k)}$. This requires a more tailored initialization of the network weights to ensure that the initial interaction potential is sufficiently short-ranged and that the corresponding graphs are not exceedingly large. To do this, we use the diffraction pattern obtained at $t=10$ to inform the initial estimate of the cutoff radius, $R^{(0)}$, which is detailed in the \hlc{Methods and Supplementary} Figure \ref{fig:figS8}. This estimate is used to pretrain the neural network such that the initial model is a decreasing function of $r$ which intercepts the $x$-axis at $R^{(0)}$. The network is subsequently trained to reproduce the simulated diffraction patterns as in the previous case study, updating all trainable parameters including $R$. Figure \ref{fig:fig3}\textbf{b} shows the true and learned cutoff functions, indicating good agreement. Note that the value of the learned parameter $R$ is not necessarily equivalent to the true $R$ in a convergent model, as it is multiplied to other trainable weights in the network. However, its value must be greater than or equal to that of the true cutoff radius in order to encompass all relevant particle interactions. The true value of $R$ can subsequently be derived from the learned cutoff function.

Figure \ref{fig:fig3}\textbf{e} shows the evolution of the predicted state of the system in real space \hlc{for one representative model and initial condition}, in good qualitative agreement with the true system state (Figure \ref{fig:fig3}\textbf{d}) even up to five times the maximum time seen during training. \hlc{We compute the mean squared error (MSE) between the centers of mass of the true and predicted states up to $t=50$ for all 10 models in the ensemble across 50 different initial configurations. These statistics appear quite similar across all models (\hlc{Supplementary} Figure \ref{fig:figS10}\textbf{a}), likely due to pretraining each model to a similar initial starting point.} Figure \ref{fig:fig3}\textbf{f} further indicates the evolution of cluster size distribution for both true (top) and predicted (bottom) states of the system. Cluster sizes are defined as the number of particles in an isolated group; specifically, this constitutes a connected graph with a maximum edge length $d$, where $d$ is approximately the center of the first peak in the distribution of all pairwise particle distances (\textit{e.g.}, \hlc{Supplementary} Figure \ref{fig:figS8}\textbf{c}). Both distributions initially exhibit a large peak at very small cluster sizes, signifying the random initial state, and evolve toward a distribution centered at larger cluster sizes.

Using the true and predicted diffraction patterns, we also compute the two-time correlation functions\hlc{, $C_2$,} at equivalent values of $Q$ indicated by the \hlc{black} dashed circle in Figure \ref{fig:fig3}\textbf{g}. Figure \ref{fig:fig3}\textbf{c} shows qualitatively similar behavior between the \hlc{$C_2$ maps} of true (\hlc{upper triangle}) and \hlc{average} predicted (\hlc{lower triangle}) dynamics \hlc{of the 10-model ensemble}. \hlc{In particular, we observe} a shift from fast to comparatively slow dynamics as the clusters \hlc{form and} stabilize, \hlc{with a temporary acceleration of the dynamics at intermediate times between $t=30$ and $t=40$}. \hlc{The lower panel of Figure \ref{fig:fig3}\textbf{c} quantifies the MAE between the true and predicted two-time correlations, showing excellent agreement until approximately $t=25$, which coincides with the period of acceleration during which relatively isolated particles in the vacant space between clusters are pulled rapidly toward a nearby, well-formed cluster. This is corroborated by the correlation times determined by fitting the one-time correlation functions extracted from the two-time maps as a function of time, plotted in \hlc{Supplementary} Figure \ref{fig:figS10}\textbf{e}. This behavior is affected by the accuracy of the cutoff radius, as slight differences can modulate the particle interactions included at the boundary and thus generate some discrepancies in the prediction. This makes the two-time correlation more sensitive to differences between the 10 models compared to the center of mass, as evidenced by the statistics of the MAE between the true and predicted $C_2$ for the same 50 initial configurations, shown in \hlc{Supplementary} Figure \ref{fig:figS10}\textbf{b}. \hlc{Supplementary} Figures \ref{fig:figS10}\textbf{c} and \textbf{d} depict the $C_2$ maps of a comparatively better and worse predicted trajectory, respectively. Their corresponding real-space analyses are shown in \hlc{Supplementary} Figure \ref{fig:figS11}. Overall, the more difficult cases exhibit shorter correlation times with more abrupt changes at intermediate times, which can be seen by comparing \hlc{Supplementary} Figures \ref{fig:figS10}\textbf{f} and \textbf{g}. Nonetheless, the overall} ability of the model to extrapolate \hlc{with qualitative and often quantitative accuracy} is a valuable source of insight into future behavior of systems with evolving rates of dynamics which may become experimentally inaccessible at long times.

\begin{figure}[t]
    \centering
    \includegraphics[width=\textwidth]{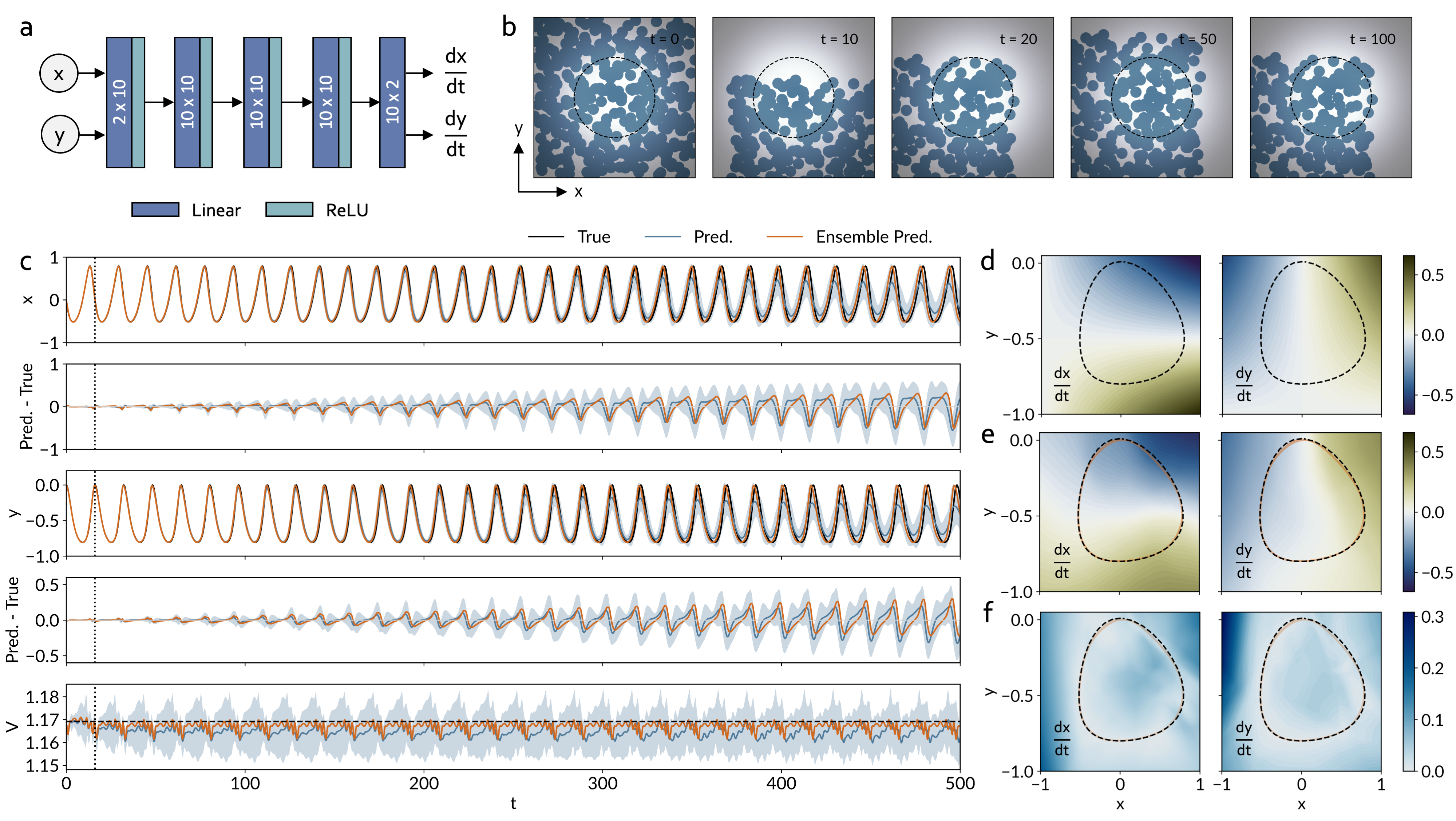}
    \caption{\textbf{Case study 3: Periodic dynamics in a fluctuating source.} \textbf{a.} Schematic of the neural network architecture modeling the fluctuation dynamics. \textbf{b.} Illustration of a time-evolving field-of-view as a result of source fluctuation. The \hlc{black} dashed circle denotes one standard deviation from the maximum of the Gaussian illumination function. \textbf{c.} \hlc{From top to bottom, the true and predicted object coordinates in the $x$ direction (row 1) and their difference (row 2); the true and predicted object coordinates in the $y$ direction (row 3) and their difference (row 4); and the value of the conserved quantity $V$ (row 5). Blue shading denotes one standard deviation across the results of the 10-model ensemble.} The dotted vertical line indicates the maximum time seen during training, \hlc{$t=16$, while the dashed black line in row 5 denotes the true value of $V$}. \textbf{d.} True governing equations of the object coordinates plotted over the domain of interest. The true object trajectory is \hlc{plotted as a black dashed line}. \textbf{e.} \hlc{Average} learned governing equations of the object coordinates \hlc{across the 10-model ensemble} plotted over the domain of interest, with the predicted object trajectory \hlc{of the ensemble model plotted in orange. The true object trajectory is replotted as a black dashed line for reference. \textbf{f.} Standard deviation of the governing equations learned by the 10 individual models. The trajectories of \textbf{d} and \textbf{e} are replotted for reference, coinciding approximately with regions of highest agreement across models.}}
    \label{fig:fig4}
\end{figure}

\paragraph{Case study 3: Fluctuating source}
In the previous case studies, the learning tasks were simplified by employing neural networks to approximate specific terms of the governing equations, such as the coupling kernel in the first example. While a larger neural network was used to approximate the cutoff function in the case of self-organizing particles, it represented only part of a larger equation with a known expected form. In the final case study, we develop a learning task in which the governing equations are modeled entirely by a neural network without prior knowledge of the functional form. For this problem, we consider a source with \hlc{positional fluctuations that is used to probe} a \hlc{stationary} test pattern formed by the random arrangement of $N=400$ particles with radii $0.07L$, where $L$ is the length of the simulation box. The source fluctuation leads to an evolving field-of-view of the object under investigation, as shown in Figure \ref{fig:fig4}\textbf{b}; thus, by recovering the apparent dynamics of the object as seen from the reference frame of the fluctuating source, the complementary source dynamics can be derived. In this case study, we define the perceived dynamics of the object to be a periodic function of its center of mass coordinates, $x(t)$ and $y(t)$, modeled after the coupled Lotka-Volterra equations \cite{wangersky1978lotka},
\begin{align}
\begin{split}
    \frac{dx}{dt} = \alpha x - \beta x y \\
    \frac{dy}{dt} = \delta x y - \gamma y,
\end{split}
\end{align}
where $\alpha$, $\beta$, $\gamma$ and $\delta$ are fixed parameters. \hlc{In our simulations, we used the set of parameters $\alpha=1/3, \beta=2/3, \gamma=1/2, \delta=1/2$, which corresponds to periodic oscillations of the coordinates with a period of $2\pi/\sqrt{\alpha\gamma} \approx 15.39$ \cite{bacaer2011lotka}. Since the Lotka-Volterra equations constitute a conservative system, the periodic solutions form closed orbits along which the conserved quantity, $V = \delta x - \gamma \ln(x) + \beta y - \alpha \ln(y)$, is constant.} We define a fully-connected neural network $\bm{F}_{\text{NN}}$ which approximates the ODE governing the evolution of $\bm{r}(t) = \left< x(t), y(t) \right >$ according to,
\begin{equation}
    \frac{d\bm{r}(t)}{dt} \approx \bm{F}_{\text{NN}}(t,\bm{r}(t);\bm{\theta}),
\end{equation}
where $\bm{\theta}$ represents the set of trainable neural network weights. Here, $\bm{F}_{\text{NN}}$ comprises an input and output layer with two neurons each, and three hidden layers each with 10 neurons (Figure \ref{fig:fig4}\textbf{a}). Each layer prior to the output is activated by a rectified linear unit (ReLU). The network is trained through comparison of predicted and calculated diffraction patterns from 100 different time series simulated up to a maximum time \hlc{$t=16$, or approximately one period,} at time intervals of $\hlc{2}$. \hlc{Each time series corresponds to different particle arrangements -- or samples -- which are generated uniformly at random within the square simulation box of length $L$. The sample centers are defined as the origin in each simulation and coincide roughly with the initial position of the probe; however, due to the random sampling, there are slight differences in the location of the sample centers and hence initial positions of the probe among the 100 different time series. During training, mini-batches of 10 time series are randomly drawn with a batch time of 16, \textit{i.e.}, the full extent of the simulation.}

The time evolution of the true and predicted coordinates \hlc{up to $t=500$} is shown in Figure \ref{fig:fig4}\textbf{c} \hlc{in terms of the average and standard deviation among the ensemble of 10 models. We also plot the value of the conserved quantity, $V$, in Figure \ref{fig:fig4}\textbf{c}. Since the neural network models are not explicitly constrained to be conservative systems, the learned trajectories can deviate slightly from the closed orbits expected analytically; as a result, the values of $V$ are only approximately level and fluctuate near the true value.} Plots of the \hlc{corresponding} true and predicted real- and reciprocal-space images are shown in \hlc{Supplementary} Figure \ref{fig:figS12}. \hlc{Overall}, the true and predicted trajectories appear to be in good agreement, \hlc{with more obvious discrepancies appearing at} approximately $t=200$. Observations at later times reveal that, \hlc{on average}, the learned period is slightly \hlc{shorter} than that of the true dynamics, resulting in a gradual divergence of true and predicted trajectories. \hlc{We can quantify the difference between these trajectories by comparing their Fourier spectra, which exhibit sharp peaks corresponding to the frequencies of oscillation. We then compute the Earth Mover's Distance (EMD), defined as the absolute difference between the cumulative distribution functions (CDFs) of the Fourier spectra, between the true and predicted trajectories from 50 different initial configurations for each of the 10 models. This is shown for three representative simulations with average, good, and poor performance across the 10 models in \hlc{Supplementary} Figures \ref{fig:figS13}\textbf{a}-\textbf{c}, whose results are visualized in Figure \ref{fig:fig4} and \hlc{Supplementary} Figures \ref{fig:figS14}\textbf{a} and \textbf{b}, respectively.}

We obtain further insight into the source of the observed discrepancy by \hlc{comparing} the learned ODEs \hlc{of each of the 10 models against the ground truth (Figure \ref{fig:fig4}\textbf{d}). In particular, \hlc{Supplementary} Figures \ref{fig:figS13}\textbf{e}-\textbf{i} show the learned ODEs of five models with varying distributions of EMD error across the 50 different simulations, while Figure \ref{fig:fig4}\textbf{e} shows the average learned ODE across the entire ensemble.} The ODEs appear to agree \hlc{most} in the vicinity of the \hlc{true} object trajectory, indicated by the \hlc{black dashed line in each of the visualizations, which is further corroborated by the plot of the standard deviation across all 10 models shown in Figure \ref{fig:fig4}\textbf{f}}. Away from the trajectory and particularly at the boundaries of the domain, the \hlc{ODEs learned by individual models do not necessarily} recover the expected behavior, likely due to the lack of observations in these regions. \hlc{By consequence, the ensemble-average shown in Figure \ref{fig:fig4}\textbf{e} resembles the ground truth ODE better than any individual model. This motivates developing an ensemble model by averaging the ODEs of the individual trained networks rather than taking the averages of their solutions, which we show in Figure \ref{fig:fig4}\textbf{c} by the orange line overlaying the true and average solutions. We also repeat the quantitative analysis in terms of Fourier spectra in \hlc{Supplementary} Figures \ref{fig:figS13}\textbf{a-d}}. In practice, \hlc{when the ground truth ODE is unknown, the analysis of learned ODEs described here is envisioned to provide both a measure of uncertainty for different regions of the domain spanned by the relevant dynamic coordinates, as well as insight into the} expected symmetries and overall complexity of the learned model, \hlc{which} could be used to place physical constraints on the network architecture or inform the use of regularization methods. While beyond the scope of the current work, the learned governing equations could further be used to guide theory development, for example by decomposition to an appropriate functional basis through methods like sparse regression.

\subsection{Measurement noise and resolution}
In practice, experimental noise reduces the quality of measured data, leading to sometimes significant differences between calculated and measured diffraction patterns. Additionally, \hlc{variations in} X-ray probe size \hlc{and coherence as well as} detector resolution can affect the clarity of reciprocal-space features. Thus, we conducted a systematic evaluation of the effects of experimental noise, probe size, \hlc{coherence,} and detector resolution on model performance. We repeated the training for the coupled moments case study across nine different levels of Poisson noise, five different values for detector resolution, $\Delta Q$, five different values for the standard deviation of the Gaussian probe, $\sigma_p$, \hlc{and five different degrees of coherence}. Poisson noise was introduced in the training data by using the calculated diffraction patterns to sample from a Poisson distribution in which the intensity acts as a proxy for the expected event frequency, $\lambda$. Specifically, we scaled the normalized intensity by a chosen factor $\lambda_{\text{max}}$, the maximum expected event frequency, and resampled each pixel value from the Poisson distribution with $\lambda$ given by the corresponding scaled intensity. To place the \hlc{absolute} intensities of true and predicted diffraction patterns on equal footing during training without explicit knowledge of $\lambda_{\text{max}}$, \hlc{we leverage the fact that at the initial condition, where the real-space solution is assumed to be known, the predicted diffraction pattern should approximate the true pattern without noise. Thus, we can rescale all diffraction patterns by the factor needed to bring the true and predicted patterns into agreement at the initial condition.} This was empirically found to perform better than normalizing all patterns by their respective maximum \hlc{or average} intensities. The model was trained for the noise-free case as well as eight different noise levels between $\lambda_{\text{max}} = 10^0$ and $\lambda_{\text{max}} = 10^7$. For $\lambda_{\text{max}} = 10^2$ through $\lambda_{\text{max}} = 10^7$, the neural ODEs were trained using a MAE loss, as was done in the perfect case. For $\lambda_{\text{max}} = 10^0$ and $\lambda_{\text{max}} = 10^1$, the Poisson statistics become significant and a Poisson negative log likelihood (NLL) was used instead.

\begin{figure}[t]
    \centering
    \includegraphics[width=\textwidth]{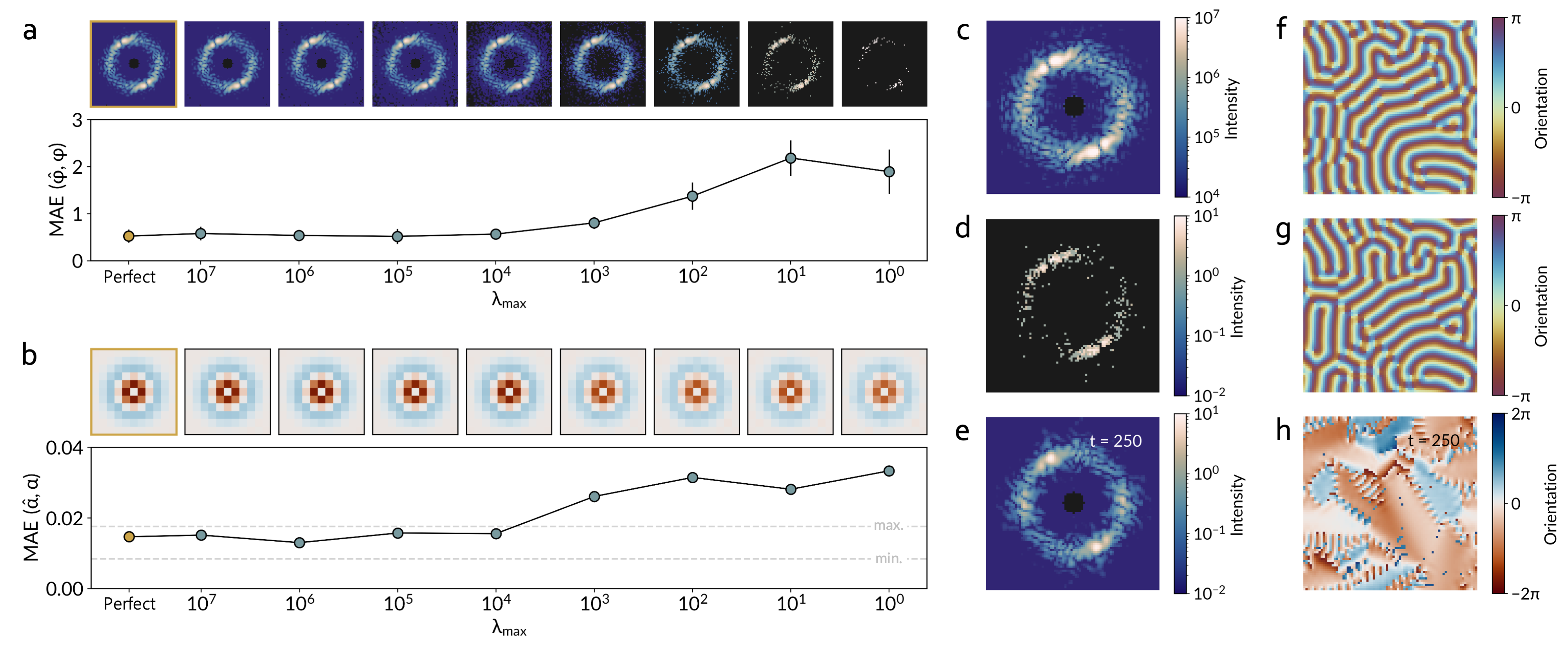}
    \caption{\textbf{Effects of measurement noise.} \textbf{a.} Mean absolute error (MAE) between the true and predicted real-space coordinates at the final extrapolated frame, $t=250$, as a function of the noise level set by $\lambda_{\text{max}}$. The results are an average over \hlc{50} different initial conditions, with error bars denoting one standard deviation. Representative diffraction patterns corresponding to each noise level are shown above. \hlc{The perfect case is distinguished by a yellow marker and a border around the representative image.} \textbf{b.} MAE between true and learned coupling kernels as a function of the noise level, $\lambda_{\text{max}}$. The learned coupling kernels corresponding to each noise level are shown above. \hlc{Gray dashed lines denote the minimum and maximum MAEs found for the 10-model ensemble under perfect conditions. The perfect case is distinguished by a yellow marker and a border around the representative image.} \textbf{c-e}. Perfect (\textbf{c}) and noisy ($\lambda_{\text{max}} = 10^1$) \textbf{d} simulated diffraction patterns at $t=250$, and the corresponding predicted diffraction pattern \textbf{e}. \textbf{f-h}. True (\textbf{f}) and predicted (\textbf{g}) states of the system in real-space corresponding to the diffraction patterns in \textbf{c} and \textbf{e}, and their difference map (\textbf{h}).}
    \label{fig:fig5}
\end{figure}

Figure \ref{fig:fig5}\textbf{a} shows the effect of increasing levels of noise (decreasing $\lambda_{\text{max}}$) on the mean absolute error between the true and predicted real-space coordinates, $\bm{\varphi}$ and $\hat{\bm{\varphi}}$, of the coupled moments at the final extrapolated frame, $t=250$. The results shown are an average over \hlc{50} different initial conditions, with error bars denoting one standard deviation. Representative diffraction patterns corresponding to each noise level are displayed above each point. In the most severe case of $\lambda_{\text{max}} = 10^0$, the average MAE is approximately 3.6 times that of the perfect case trained without noise. Similarly, we visualize the learned coupling kernels and plot the errors as a function of $\lambda_{\text{max}}$ in Figure \ref{fig:fig5}\textbf{b}. \hlc{For $\lambda_{\text{max}} > 10^3$, the MAE falls within the maximum error found for the 10-model ensemble under perfect conditions, suggesting robustness to noise until approximately $\lambda_{\text{max}} = 10^3$.} While the learned coupling kernel is less well-resolved at higher noise levels, the anticipated symmetry and overall pattern is still evident. As an example, we show representative images of true and predicted real- and reciprocal-space patterns for the case of $\lambda_{\text{max}} = 10^1$ in detail in Figures \ref{fig:fig5}\textbf{c} - \ref{fig:fig5}\textbf{h}. Additional images for all simulated values of $\lambda_{\text{max}}$ are reported in \hlc{Supplementary} Figure \ref{fig:figS15}.

\begin{figure}[t]
    \centering
    \includegraphics[width=\textwidth]{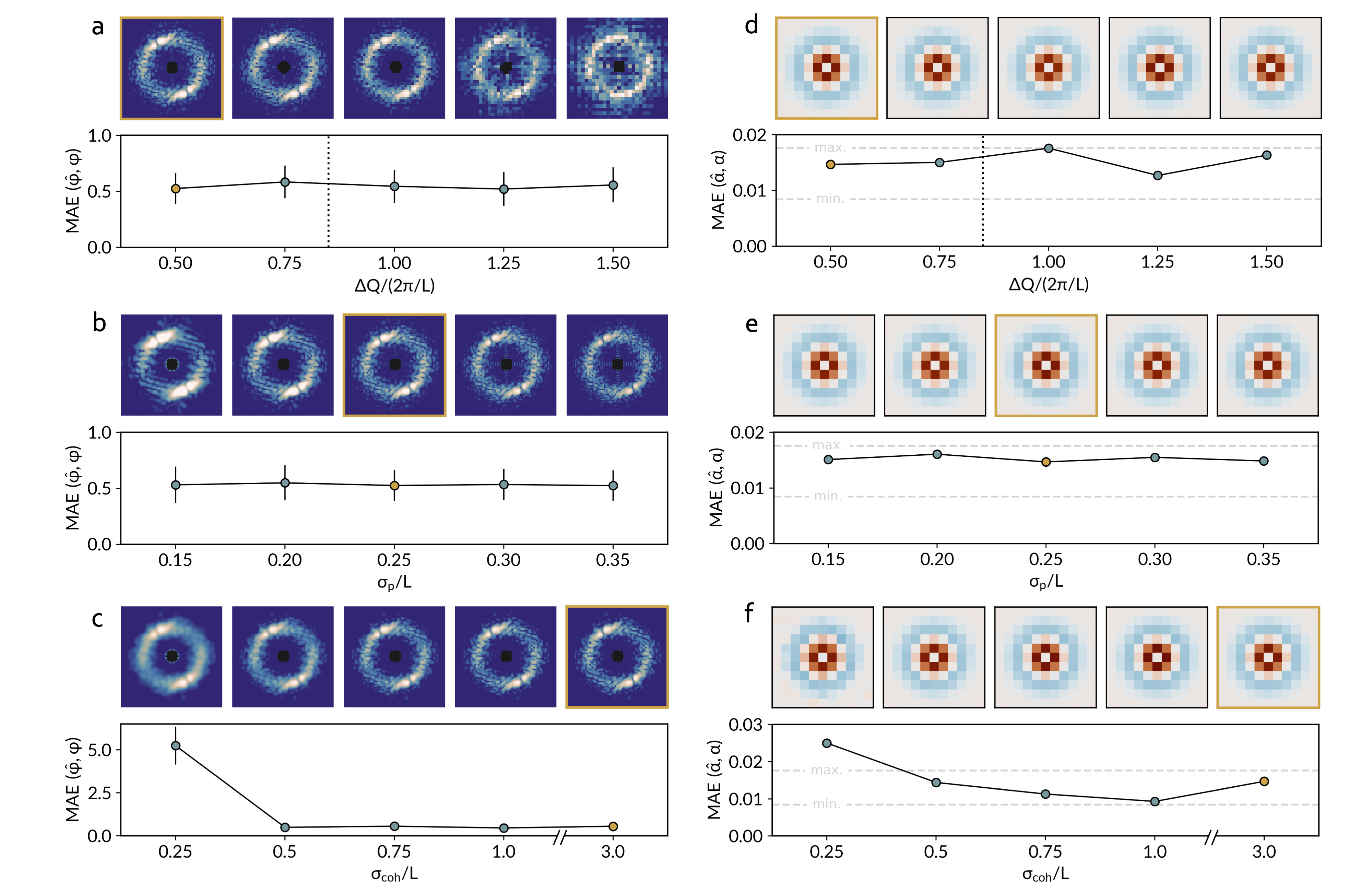}
    \caption{\textbf{Effects of \hlc{detector} resolution, probe size, \hlc{and coherence}.} \hlc{\textbf{a-c.} Mean absolute error (MAE) between the true and predicted real-space coordinates at the final extrapolated frame, $t=250$, as a function of detector resolution (\textbf{a}), probe size (\textbf{b}), and coherence (\textbf{c}). The results are an average over 50 different initial conditions, with error bars denoting one standard deviation. Representative diffraction patterns are displayed above each condition. The original conditions used for this case study are distinguished by yellow markers and borders around representative images. The black dotted line in \textbf{a} denotes the value of $\Delta Q$ at the Nyquist-Shannon sampling criterion. \textbf{d-f.} MAE between the true and learned coupling kernels as a function of detector resolution (\textbf{d}), probe size (\textbf{e}), and coherence (\textbf{f}). Images of the learned kernels are displayed above each condition. The original conditions used for this case study are distinguished by yellow markers and borders around representative images. Gray dashed lines denote the minimum and maximum MAEs found for the 10-model ensemble under perfect conditions. The black dotted line in \textbf{d} denotes the value of $\Delta Q$ at the Nyquist-Shannon sampling criterion.}}
    \label{fig:fig6}
\end{figure}

Finally, we also report the performance of the proposed framework on detector resolution, probe size, \hlc{and coherence}. \hlc{Detector resolution was varied from $\Delta Q = \pi/L$ to $\Delta Q = 3\pi/L$. We note that, for the probe size of $\sigma_p = L/4$ used in these examples, the last 3 values of $\Delta Q$ fall outside of the Nyquist-Shannon sampling criterion, which requires that $\Delta Q \leq 2 \pi/W$, where $W$ is the transverse size of a finite object or alternately of the illuminated area \cite{van2004coherent, prosekov2021methods}. In our example, $W$ was defined as the full width at half maximum (FWHM) of the Gaussian probe. Examples with $\Delta Q > 2 \pi/W$ could not be unambiguously inverted to the corresponding real-space images, further motivating use of the proposed framework. When varying probe size, $\Delta Q$ was maintained at $\pi/L$ while $\sigma_p$ was varied between $0.15L$ and $0.35L$. Finally, we vary the degree of coherence by convolving the scattered intensity with the Fourier transform of a Gaussian mutual coherence function (MCF) with a standard deviation $\sigma_{coh}$ \cite{kim2022performance}, which we varied from fully coherent ($\sigma_{coh} \geq 3L$) to $\sigma_{coh} = L/4$.} Figures \ref{fig:fig6}\textbf{a}\hlc{-\textbf{c}} show representative diffraction patterns varying the resolution\hlc{, probe size, and coherence,} respectively. In \hlc{nearly all} cases, we found that the average MAE between the true and predicted real-space coordinates at the final extrapolated frame \hlc{across 50 different initial configurations} was not statistically different \hlc{from that under the original conditions used for this case study. The only exception was the case of $\sigma_{coh} = L/4$, for which the error was approximately 10 times that of the perfectly coherent case.} While there appears to be some variation in the errors between learned and true kernels \hlc{under different conditions}, it is still well below the discrepancy seen between the kernels of very noisy ($\lambda_{\text{max}} > 10^2$) and noise-free examples and is most likely the result of differences in the models' weight initialization and training history. \hlc{For reference, we indicate the minimum and maximum MAEs found for the 10-model ensemble under the original conditions used for this case study in each plot; all MAEs are found to be within this range, except that at a partial coherence of $\sigma_{coh} = L/4$.} As a result, we conclude that the model's performance is relatively insensitive to \hlc{most relevant conditions for} detector resolution, probe size, \hlc{and coherence that may be encountered experimentally}.

\subsection{Experimental proof-of-concept}
\begin{figure}[t]
    \centering
    \includegraphics[width=\textwidth]{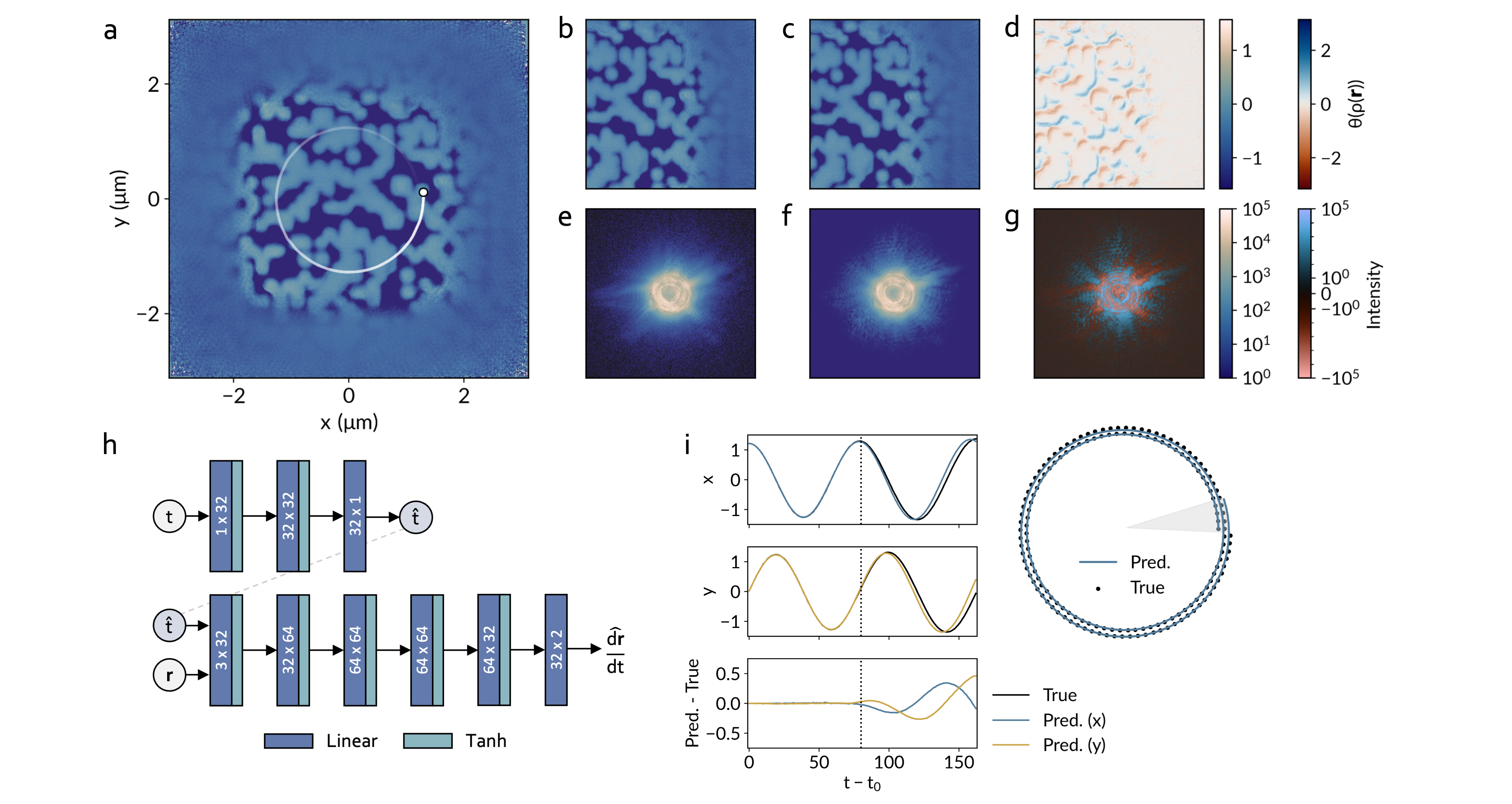}
    \caption{\textbf{Experimental proof-of-concept.} \textbf{a.} The real-space phase of the test pattern. The probe trajectory is illustrated in white up to the time $t-t_0=80$, denoted by the white circle. \textbf{b.} The real-space frame centered at the true scan point at $t-t_0=80$, indicated by the white circle in \textbf{a}. \textbf{c.} The real-space frame centered at the predicted scan point at $t-t_0=80$. \textbf{d.} The difference map between \textbf{b} and \textbf{c}. \textbf{e.} Experimentally-measured coherent diffraction pattern corresponding to the real-space image in \textbf{b}. \textbf{f.} The coherent diffraction pattern calculated from the real-space image in \textbf{c}. \textbf{g.} The difference map between \textbf{e} and \textbf{f}. \textbf{h.} Schematic of the neural network architecture. \textbf{i.} The true and predicted probe coordinates in horizontal (top row) and vertical (middle row) directions. The difference between true and predicted coordinates is plotted in the bottom row. The dotted vertical line indicates the maximum time seen during training, $t-t_0=80$. \hlc{The accumulated error in the predicted path is quantified in terms of the angular lag at the final extrapolation point, approximately $0.11\pi$, illustrated by the light gray wedge on the adjacent visualization of the true and predicted trajectories.}}
    \label{fig:fig7}
\end{figure}

To investigate the performance of the proposed framework in an experimental context, we develop a simple proof-of-concept relying on experimental diffraction data for learning. In particular, we considered the problem of recovering the trajectory of an X-ray beam scanned across an object along a predefined path from the resulting coherent diffraction data. To do this, we used X-ray ptychographic data obtained from a tungsten test pattern as reported in Ref. \citenum{cherukara2020ai}. In this experiment, coherent diffraction patterns were obtained with 0.3s acquisitions over a regular grid of scan points across a tungsten test pattern etched with random features \hlc{(Supplementary Figure \ref{fig:figS17}\textbf{a})}. The real-space amplitude and phase of the object as well as the complex probe were then recovered using an iterative phase retrieval algorithm. We refer to Ref. \citenum{cherukara2020ai} for additional details about the experimental dataset. To define the learning task for this case study, we considered a continuous trajectory through the scan points given by,
\begin{align}
\begin{split}
    \frac{dx}{dt} &= \frac{1}{2\sqrt{t}}\left( \frac{x}{\sqrt{t}} - \omega y \right) \\
    \frac{dy}{dt} &= \frac{1}{2\sqrt{t}}\left( \frac{y}{\sqrt{t}} + \omega x \right),
\end{split}
\end{align}
where $x(t)$ and $y(t)$ are coordinates of the center of the incident beam in the plane of the object, and \hlc{$\omega=4$} is a constant parameter. These equations describe a circular spiral centered on the test pattern. For the training data, we isolated a segment of this trajectory, shown in Figure \ref{fig:fig7}\textbf{a}, comprising 80 sequential scan points starting at a time $t_0 > 0$ \hlc{which approximately span one revolution}. At each scan point along the trajectory, a diffraction pattern of $512 \times 512$ pixels is obtained as the square modulus of the Fourier transform of the illuminated real-space object multiplied by the complex probe. A representative real- and reciprocal- space pair is shown in Figures \ref{fig:fig7}\textbf{b} and \textbf{e}. \hlc{The learning task in this proof-of-concept resembles that of the fluctuating source case study; namely, recovering two degrees of freedom specifying the motion of the probe over a stationary sample. However, apart from differences in the dynamical equations, the proof-of-concept experiment also differs from the final case study in that the beam is focused by a Fresnel zone plate and the sample can be regarded as effectively infinite.} To predict the trajectory from experimental data, we define the neural network $\bm{F}_{\text{NN}}$ which approximates the ODE governing the evolution of $\bm{r}(t) = \left< x(t), y(t) \right >$ according to,
\begin{equation}
    \frac{d\bm{r}(t)}{dt} \approx \bm{F}_{\text{NN}}(t,\bm{r}(t);\bm{\theta}),
\end{equation}
where $\bm{\theta}$ represents the set of trainable neural network weights. The network architecture of $\bm{F}_{\text{NN}}$ is shown in Figure \ref{fig:fig7}\textbf{h}. Due to the disparate scales of the spatial and time coordinates, a shallow network is first used to featurize $t$, which is then passed to a deeper network along with the current value of $\bm{r}(t)$ to predict $d\bm{r}/dt$. To extract the appropriate frames centered at arbitrary values of $\bm{r}(t)$, an interpolating function of the full real-space image was first calculated using a regular grid interpolant and could then be sampled at arbitrary values of $\bm{r}(t)$ predicted by the network. Figure \ref{fig:fig7}\textbf{c} shows the real-space image at the predicted scan location for the same time point shown in Figure \ref{fig:fig7}\textbf{b}, along with the calculated diffraction pattern (Figure \ref{fig:fig7}\textbf{f}). As in the computational case studies, true and predicted diffraction patterns were compared using their mean absolute error in order to train the network. The network was trained through comparison of predicted and experimental diffraction patterns of 80 sequential time points which capture approximately one periodic cycle, \hlc{using mini-batches of size 5 with a batch time of 35 time points}. The time evolution of the true and predicted coordinates of the trained network are shown in Figure \ref{fig:fig7}\textbf{i} up to twice the maximum time observed during training, at which point the trajectories are seen to gradually diverge. \hlc{This results from a slight dilation of the predicted path compared to the true, which is illustrated in the visualization of the true and predicted trajectories in Figure \ref{fig:fig7}\textbf{i}. We quantify the accumulated error in the predicted path in terms of the angular lag at the final extrapolation point, which is about $0.11\pi$. Due to the quasi-periodic nature of the motion, we also include a comparison of the true and predicted Fourier spectra, as performed in the final computational case study, in \hlc{Supplementary} Figure \ref{fig:figS17}\textbf{b}. Nonetheless,} the qualitative behavior at long times remains comparable between true and predicted trajectories. Additional real- and reciprocal-space images along the trajectory are reported in \hlc{Supplementary} Figure \ref{fig:figS18}. While simple, this learning task serves as a valuable proof-of-concept that the proposed framework is transferable to experimental datasets and is a helpful reference point for assessing performance on more complicated systems with unknown dynamics.

\section{Discussion}
As the framework developed in this work is envisioned as a general tool for extracting dynamics from coherent scattering measurements, here we emphasize several practical considerations for its use. First, it is important to recognize that the learned equations generalize only to different initial conditions of the dynamic system on which the model was trained. Therefore, \hlc{a} key use case of a trained model \hlc{is} expected to be the extrapolation of dynamics beyond experimentally-accessible observations, which can inform the design and planning of experiments. \hlc{We note that the case studies explored in this work generally approach a more static state or are in a fluctuating equilibrium; as a result, capturing dynamics that explicitly accelerate with time may require further development of the method. Additional applications of the trained model include} computational investigation of dynamics proceeding from different initial conditions from those available at training time, and guidance toward the development of theoretical models of complex dynamics through analysis of the learned ODE.

Second, accurate information about the real-space object and relevant dynamic coordinates is needed at the initial conditions. This generally assumes that longer \hlc{acquisitions} of the system under relatively static conditions can be conducted and used to recover the real-space image at the initial state, for instance through traditional phase retrieval methods. \hlc{As an example, we propose that certain emergent dynamics, which are tunable by external stimuli such as light \cite{zinn2022emergent}, temperature, or electric or magnetic fields \cite{grigoriew2010dynamic} and drive the system toward a new phase or state of equilibrium, could be suitable first-experiments for validating the proposed framework by enabling comparatively static scans to be captured at the initial conditions followed by fast measurements of subsequent dynamics.}

Finally, there is an inherent trade-off between problem complexity and the dimensionality of the selected dynamic coordinates, which is often determined by available prior knowledge about the scattering medium. For instance, in the case study of coupled moments, the diffracted intensity was modulated by the changing orientation of moments via an effective form factor while keeping moment positions fixed. In the remaining two case studies, changes in object positions relative to the incident beam governed intensity fluctuations while form factors were fixed functions of $Q$. Generalizations of these learning tasks with fewer prior assumptions are possible but were not explored in the present work. For instance, the form factors in the latter two case studies could be implemented as trainable functions given an informed initial guess, \textit{e.g.}, a spherical form factor of an estimated particle size used to initialize the approximation of a more irregularly shaped scatterer. Even more generally, individual pixels of the real-space image can be considered dynamic coordinates provided a sufficiently large neural network approximator. Though not explored in the present work, these reformulations of the learning tasks can be readily performed within the proposed framework and may be necessary when working with experimental samples that are less well understood.

In conclusion, we demonstrated the application of a scientific machine learning framework that enables the extraction of physical insights about real-space dynamics directly from time-resolved coherent scattering measurements. This framework parameterizes unknown real-space dynamics using neural differential equations of the system's dynamic coordinates, which are optimized by comparing the time series of true and predicted coherent diffraction patterns. Real-space information is needed only at the initial conditions, making it possible to perform high-fidelity measurements only under static conditions and fast scans thereafter to enable inference of subsequent dynamics. The performance of the framework is evaluated in the context of three different computational systems and an experimental proof-of-concept, demonstrating its versatility and potential applications. In practice, this framework is envisioned to assist and accelerate data interpretation and theory development from coherent scattering in a broad range of materials systems, thereby advancing our understanding of dynamic phenomena at the mesoscale.

\section*{Methods}
\subsection*{Coherent scattering calculations}
We compute the elastic scattered intensity from a coherent X-ray probe incident on the scattering medium with predicted coordinates $\bm{\hat{x}}(t;\bm{\theta})$ as \cite{mohanty2022computational},
\begin{equation}
    \hat{I}(\bm{Q},t;\bm{\theta}) = \left|\sum_{j=1}^N \hlc{P(\bm{Q})*} f_j(\bm{Q},\bm{\hat{x}}(t;\bm{\theta}))e^{i\bm{Q} \cdot \bm{r}_j(\bm{\hat{x}}(t;\bm{\theta}))}\right|^2,
\end{equation}
where $\bm{Q}$ denotes the wavevector change, and the scattering medium, \textit{i.e.}, real-space object, comprises $N$ scatterers with positions $\bm{r}_j(\bm{\hat{x}}(t;\bm{\theta}))$ and form factors $f_j(\bm{Q},\bm{\hat{x}}(t;\bm{\theta}))$, each with potential dependence on the dynamic coordinates of interest. \hlc{$P(\bm{Q})$ is the Fourier transform of the X-ray probe function, and $*$ denotes convolution.} We model the illumination of the scattering medium by the X-ray probe as a two-dimensional Gaussian. This is incorporated by convolving the Fourier transform of the object with that of the Gaussian illumination function prior to computing its square modulus. Typically, the bright, unscattered beam at the center is obstructed to protect the detector and was therefore also masked in the intensity calculations performed here. As long as the scattering forward model remains a differentiable function of the inputs, additional refinement to the calculation can be incorporated to better reflect experimental conditions. \hlc{Below, we write the explicit intensity calculations used in each of the three computational case studies.}

\paragraph{\hlc{Case study 1: Locally-coupled moments}}
\hlc{In the case of locally-coupled moments, the dynamic coordinates were the moment orientations $\varphi_j(t)$ relative to the incident beam. All moments were considered spatially fixed on a square lattice at positions $\bm{r}_j$. To capture the time-evolution of the moment orientations, the form factor was modeled as a function of $\varphi_j(t)$ and governed the time-evolution of the scattered intensity. To maintain generality, the total scattered intensity was modeled as the incoherent sum of $\varphi$-independent and -dependent contributions, $I^{(0)}$ and $I^{(1)}$, respectively, \textit{i.e.},
\begin{equation}
    I(\bm{Q},t) = I^{(0)}(\bm{Q}) + I^{(1)}(\bm{Q},t).
\end{equation}
As this system exhibited no time-dependent evolution of $\bm{r}_j$, the first term was a radially-symmetric function of $\bm{Q}$ for all $t$ given by,
\begin{equation}
    I^{(0)}(\bm{Q}) = \left|P(\bm{Q}) * \sum_{j=1}^N f^{(0)}e^{i\bm{Q} \cdot \bm{r}_j}\right|^2,
\end{equation}
where $f^{(0)}$ is constant. The second term was given by,
\begin{equation}
    I^{(1)}(\bm{Q},t) = \left|P(\bm{Q}) * \sum_{j=1}^N f^{(1)}(\varphi_j(t))e^{i\bm{Q} \cdot \bm{r}_j}\right|^2,
\end{equation}
where $f^{(1)}(\varphi_j(t)) = \cos{(\varphi_j(t))}$. In \hlc{Supplementary} Figure \ref{fig:figS1}, we visualize the key quantities relevant to the scattering calculation at a given time point. The left panel shows the real-space object $\mathcal{O}(\bm{r})$ overlaid with the Gaussian illumination function $p(\bm{r})$. The middle panel shows the convolution kernel corresponding to the truncated Fourier transform of the Gaussian probe, $P(\bm{Q}) = \mathcal{F}\{p(\bm{r})\}$, where $\mathcal{F}$ denotes the Fourier transform. Finally, the right panel shows $I(\bm{Q})$ for the corresponding $\mathcal{O}(\bm{r})$ and $P(\bm{Q})$.}

\paragraph{\hlc{Case study 2: Self-organizing particles}}
\hlc{In the case of self-organizing particles, the dynamic coordinates were the particle positions $\bm{r}_j(t)$. As all particles were considered to be identical, the same form factor \hlc{amplitude} $f_s(\bm{Q})$ was used for each particle, given by,
\begin{equation}
    \hlc{f_s(\bm{Q}) = \frac{3}{(Qa)^3} \left(\sin{(Qa)} - Qa\cos{(Qa)}\right)},
\end{equation}
where $Q = |\bm{Q}|$ and $a$ is the particle radius. The final intensity was then calculated as,
\begin{equation}
    I(\bm{Q},t) = \left|P(\bm{Q}) * \sum_{j=1}^N f_s(\bm{Q})e^{i\bm{Q} \cdot \bm{r}_j(t)}\right|^2.
\end{equation}}

\paragraph{\hlc{Case study 3: Fluctuating source}}
\hlc{In the case of the fluctuating source, the dynamic coordinates were the center of mass coordinates of the object under observation, $\bm{r}(t) = \left<x(t), y(t)\right>$. Since the object consisted of $N$ identical particles with fixed relative positions, the center of mass coordinates were related to the instantaneous particle locations through,
\begin{equation}
    \bm{r}(t) = \frac{1}{N}\sum_{j=1}^N \bm{r}_j(t) \\
\end{equation}
All particles were considered to be identical, and hence the same form factor $f_s(\bm{Q})$ was used for each particle as in the case of self-organizing particles. The intensity was then calculated in the same manner according to,
\begin{equation}
    I(\bm{Q},t) = \left|P(\bm{Q}) * \sum_{j=1}^N f_s(\bm{Q})e^{i\bm{Q} \cdot \bm{r}_j(t)}\right|^2.
\end{equation}}

\subsection*{Machine learning}
We implement this framework using \texttt{PyTorch} \cite{paszke2017automatic} and employ the \texttt{torchdiffeq} \cite{torchdiffeq} library of ODE solvers for numerical integration, which supports backpropagation through ODE solutions using the adjoint sensitivity method. Neural networks were trained using the Adam optimizer, with mini-batching performed both on realizations of the dynamical system (\textit{i.e.}, initial conditions) and in time. The initial learning rates were empirically set to $5 \times 10^{-3}$, $10^{-3}$, and $10^{-2}$ for the three computational case studies, respectively, and to $3 \times 10^{-4}$ for the experimental proof-of-concept. In all cases, learning rates were reduced by a factor of 2 over the course of training if the loss \hlc{increased by more than twice} the running average value. Training concluded when the loss no longer improved appreciably compared to batch-to-batch fluctuations.

\subsection*{\hlc{Two-time correlation function calculations}}
\hlc{We calculate the two-time correlation function as the autocovariance of the intensity normalized by its standard deviation \cite{brown1997speckle,bikondoa2017use},
\begin{equation}
    \hlc{C_2(Q,t_1,t_2) = \frac{\left< I(Q,t_1) I(Q,t_2) \right> - \left< I(Q,t_1) \right> \left< I(Q,t_2) \right>}{\sqrt{\left< I(Q,t_1)^2 \right> - \left< I(Q,t_1) \right>^2} \sqrt{\left< I(Q,t_2)^2 \right> - \left< I(Q,t_2) \right>^2}},}
\end{equation}
where $I(Q,t_1)$ and $I(Q,t_2)$ are intensities at times $t_1$ and $t_2$ at a scattering wavevector of magnitude $Q$, and $\left< \cdot \right>$ denotes an average over all equivalent $Q$. When extracting one-time correlation functions from the two-time maps, the delay time direction is taken along lines perpendicular to the $t_1 = t_2$ diagonal, which is considered the observation time direction, and the delay time magnitude is given by $\Delta t = |t_2 - t_1|$.}

\subsection*{\hlc{Pretraining details for Case study 2}}
\hlc{As noted in the main text, proper initialization of the network describing pairwise particle interactions is necessary to ensure that the initial interaction potential is sufficiently short-ranged. To inform this initialization without detailed knowledge of the underlying system, we outline and validate a simple procedure based only on the available diffraction patterns of the evolving system. In particular, as the system approaches a clustered state, we observe the emergence of a bright ring of diffraction spots close to the central peak, highlighted in \hlc{Supplementary} Figure \ref{fig:figS8}\textbf{a}. Such systematic modulations of intensity are often indicative of structural symmetries in real-space; in fact, the bright features appear at a length scale comparable to the average spacing between neighboring clusters (\hlc{Supplementary} Figure \ref{fig:figS8}\textbf{b}). By histogramming all pairwise particle distances in the final simulation frame (\hlc{Supplementary} Figure \ref{fig:figS8}\textbf{c}), we find that this length scale corresponds closely with the second peak of the resulting distribution, \textit{i.e.}, the average distance between neighboring clusters. (The first peak near 0 corresponds to pairwise distances between particles of the same cluster.) This suggests that the length scale corresponding to the emergent features is a reasonable initial choice for the initial cutoff radius, $R^{(0)}$. This estimate is used to pretrain the neural network such that the initial model is a decreasing function of $r$ which intercepts the $x$-axis at $R^{(0)}$. Specifically, we pretrain $F^{(0)}_{\text{NN}}$ to approximate, 
\begin{equation}
    F^{(0)}_{\text{NN}}(t,r; R,\bm{\theta}) \approx \frac{1}{\hlc{n}}\frac{R^{(0)}-r}{r}.
\end{equation}
\hlc{Supplementary} Figure \ref{fig:figS8}\textbf{d} plots the predicted cutoff function before (left) and after (right) pretraining. This ensures that the initial interaction potential is short-ranged and therefore suitable for constructing graphs based on pairwise particle interactions. \hlc{As expected}, the pretrained network is still far from the target solution, as illustrated in \hlc{Supplementary} Figures \ref{fig:figS8}\textbf{e} and \textbf{f}. \hlc{Supplementary} Figure \ref{fig:figS8}\textbf{f} \hlc{shows} the time evolution of the predicted states of the system at selected time points up to $t = 50$, which do not match the expected dynamics (\hlc{Supplementary} Figure \ref{fig:figS8}\textbf{e}). This is further reflected in the discrepancy between true and predicted cluster size distributions, shown in \hlc{Supplementary} Figure \ref{fig:figS8}\textbf{g}. \hlc{The pretrained model greatly overestimates cluster spacing because the initial estimate of the cutoff function is highly approximate; its purpose is only to reasonably restrict the size of the graphs to exclude unnecessarily long-ranged interactions.} We also observe significant differences between the true and predicted diffraction patterns at the final state in \hlc{Supplementary} Figures \ref{fig:figS8}\textbf{h}-\textbf{j}. Thus, it is necessary to further refine the cutoff function in order to obtain the optimized results shown in Figure \ref{fig:fig3}, which is accomplished by \hlc{further} training the network following the proposed framework.}

\section*{Data availability}
The data used in this study are made available at \texttt{https://zenodo.org/records/10204977}.

\section*{Code availability}
The code used in this study is made available at \texttt{https://github.com/ninarina12/dynamiCXS}.

\section*{Acknowledgments}
N.A. would like to thank Zhantao Chen and Saugat Kandel for their helpful discussions and insights on this research. This material is based upon work supported by Laboratory Directed Research and Development (LDRD) funding from Argonne National Laboratory, provided by the Director, Office of Science, of the U.S. Department of Energy under Contract No. DE-AC02-06CH11357. M.K.Y.C. acknowledges the support from the BES SUFD Early Career award. Work performed at the Center for Nanoscale Materials, a U.S. Department of Energy Office of Science User Facility, was supported by the U.S. DOE, Office of Basic Energy Sciences, under Contract No. DE-AC02-06CH11357. This research used resources of the Advanced Photon Source, a U.S. Department of Energy (DOE) Office of Science User Facility operated for the DOE Office of Science by Argonne National Laboratory under Contract No. DE-AC02-06CH11357. 

\section*{\hlc{Author contributions}}
\hlc{N.A. conceived the presented study with support from M.J.C. and M.K.Y.C. N.A. developed the codebase and performed the numerical and machine learning computations. T.Z carried out the original experiment and prepared and analyzed the experimental data. N.A. and Q.Z. devised the experimental proof-of-concept. S.N., M.J.C., and M.K.Y.C. provided critical feedback and helped supervise the project. N.A. wrote the manuscript with input from all authors. All authors have read and approved the manuscript.}

\section*{Competing interests}
The authors declare no competing interests.

\section*{Additional information}
Supplemental Material is available for this paper. Correspondence and requests for materials should be addressed to Nina Andrejevic (\texttt{nandrejevic@anl.gov}), Maria K. Y. Chan (\texttt{mchan@anl.gov}), or Mathew J. Cherukara (\texttt{mcherukara@anl.gov}).

\bibliographystyle{unsrt}
\bibliography{main}

\clearpage

\input{supp}
\begin{figure}[h]
    \centering
    \includegraphics[width=\textwidth]{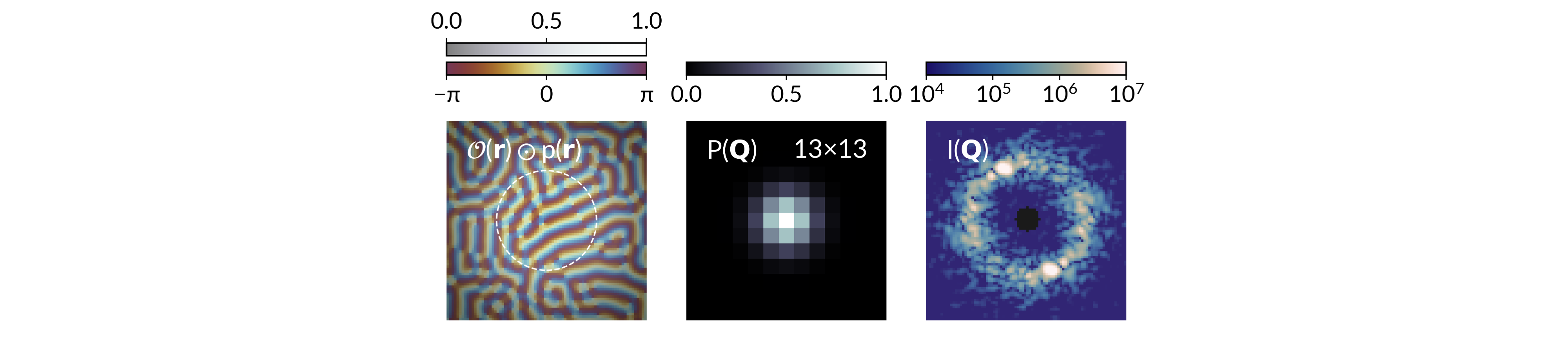}
    \caption{\textbf{Visualizing calculated quantities.} (Left) The real-space object $\mathcal{O}(\bm{r})$ overlaid with the Gaussian illumination function $p(\bm{r})$. (Middle) The convolution kernel corresponding to the truncated Fourier transform of the Gaussian probe, $P(\bm{Q})$. (Right) $I(\bm{Q})$ for the corresponding $\mathcal{O}(\bm{r})$ and $P(\bm{Q})$. The bright central peak is masked.}
    \label{fig:figS1}
\end{figure}

\begin{figure}[bh]
    \centering
    \includegraphics[width=\textwidth]{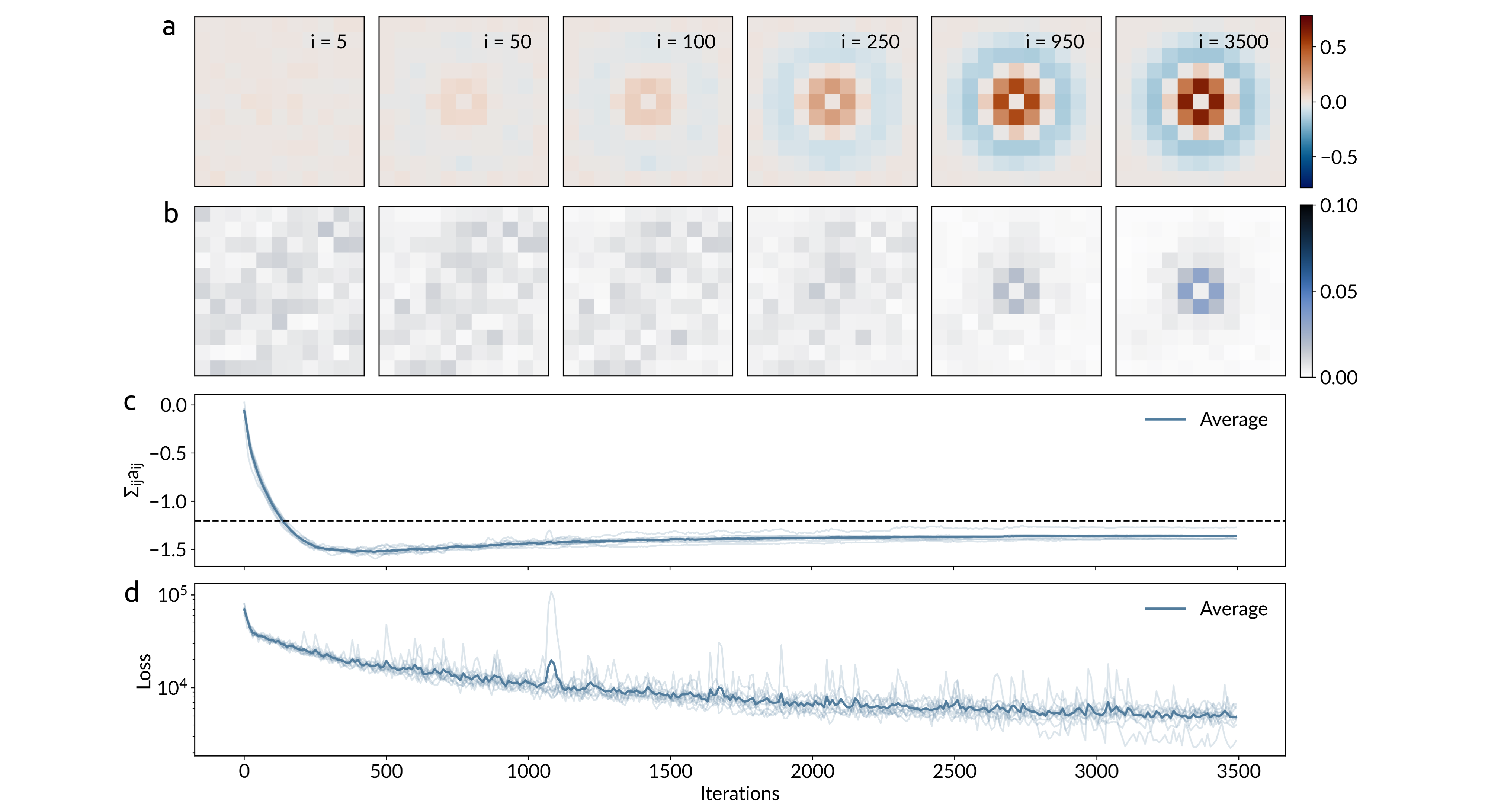}
    \caption{\hlc{\textbf{Training history for a 10-model ensemble.} \textbf{a.} Evolution of the average learned spatial coupling kernel from an ensemble of 10 models over the course of training, depicted after 5, 50, 100, 250, 950, and 3500 iterations. \textbf{b.} The evolution of the corresponding standard deviation of the coupling kernels in \textbf{a}. \textbf{c.} The sum over the kernel values $a_{ij}$ as a function of training time for each of the 10 models (light blue lines) and their average (bold blue line). The dashed black line denotes the sum over values of the ground truth kernel. \textbf{d.} The training loss, computed as the MAE between the true and predicted diffraction patterns, as a function of training time for each of the 10 models (light blue lines) and their average (bold blue line).}}
    \label{fig:figS2}
\end{figure}

\begin{figure}[t]
    \centering
    \includegraphics[width=\textwidth]{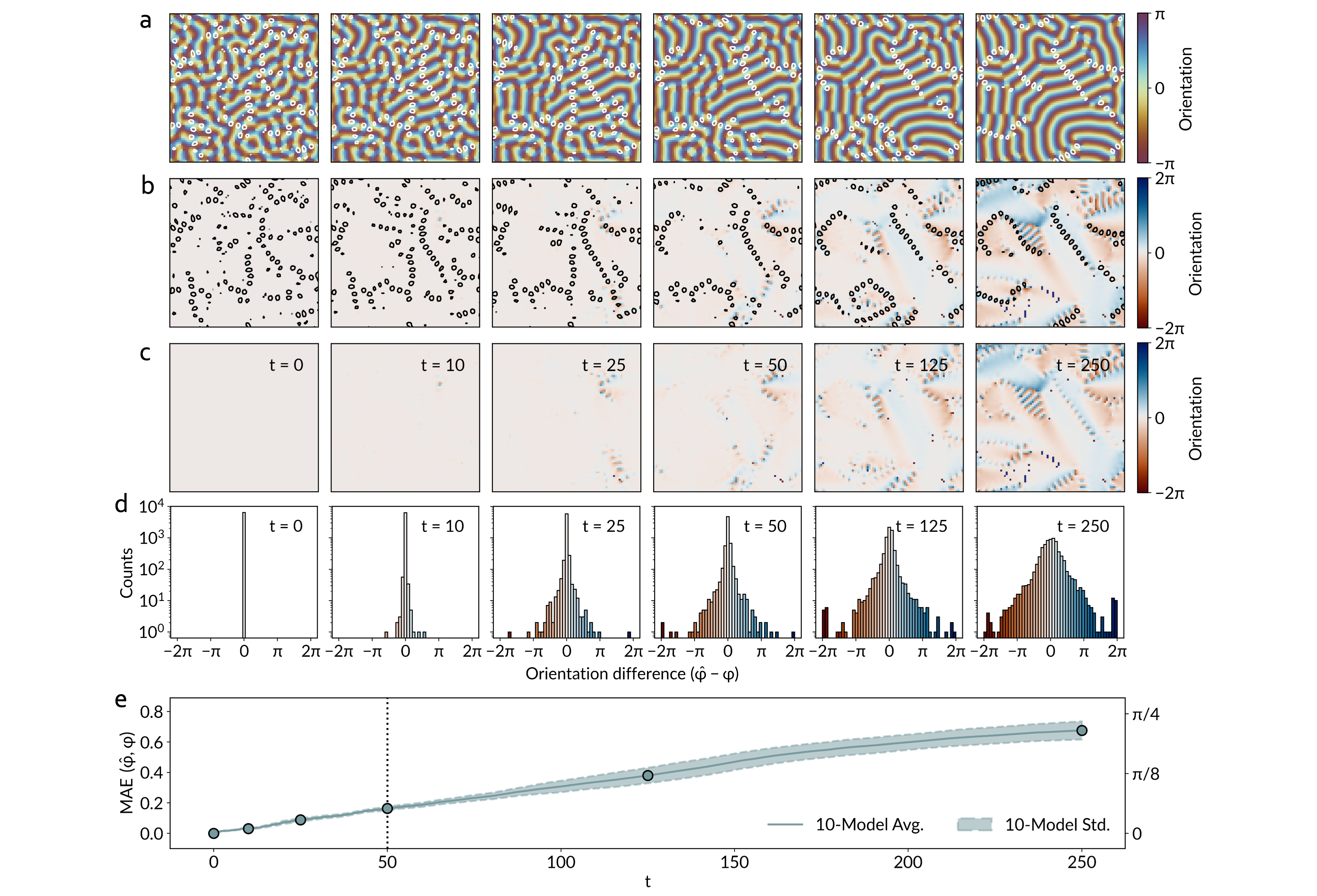}
    \caption{\hlc{\textbf{Quantification and interpretation of real-space errors.} \textbf{a.} The time evolution of the true state of the system at selected time points overlaid by white contours encircling points of high local vorticity in the moments. \textbf{b-c.} The evolution of the difference map between the predicted and true system state in \textbf{a}, with (\textbf{b}) and without (\textbf{c}) the vorticity contours displayed. It can be seen that errors in the prediction tend to originate and propagate along paths of high vorticity. \textbf{d.} The distribution of errors within each $80 \times 80$ frame of moments in \textbf{c}. \textbf{e.} The average MAE between the true and predicted real-space images for a 10-model ensemble as a function of time. Shaded area denotes one standard deviation. Note that the right axis reports the error in terms of the average difference in moment orientation, which is just under $\pi/4$ or $12.5\%$ at the maximal observation point. The black dashed line indicates the maximum time seen during training.}}
    \label{fig:figS3}
\end{figure}

\begin{figure}[t]
    \centering
    \includegraphics[width=\textwidth]{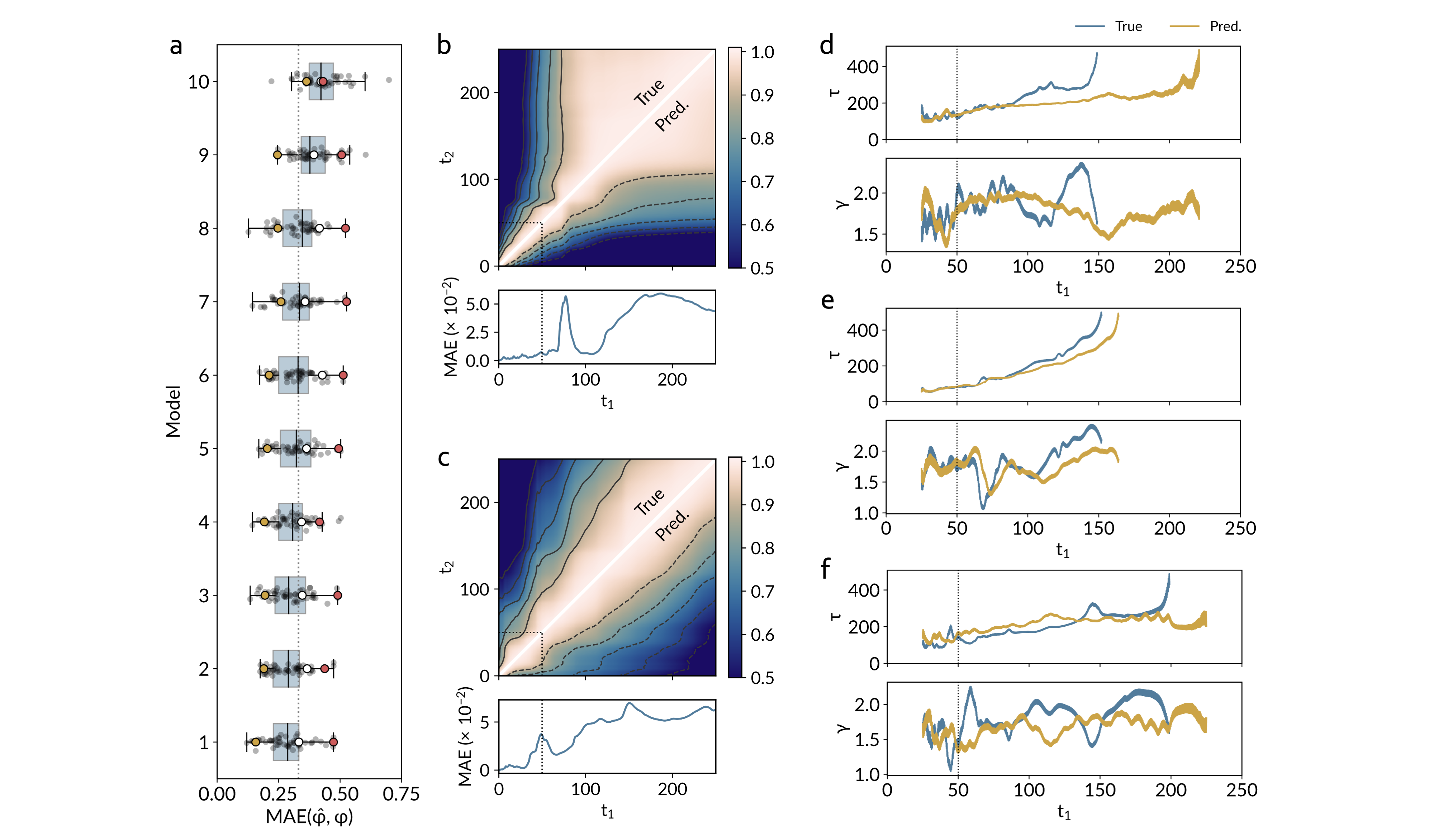}
    \caption{\hlc{\textbf{Two-time correlation function analysis.} \textbf{a.} Boxplots of the MAE between true and predicted real-space images for each of the 10 models in the ensemble for 50 different initial configurations of the system. Light gray points correspond to each individual data point (initial configuration). Three highlighted data points in each boxplot correspond to the three examples visualized in Figure \ref{fig:fig2} (white), Figure \ref{fig:figS5} (yellow), and Figure \ref{fig:figS6} (red). \textbf{b-c.} The two-time intensity-intensity correlation functions calculated from the diffraction patterns of the true (upper triangle) and average predicted (lower triangle) systems for the initial configurations given by the yellow (\textbf{b}) and red (\textbf{c}) points in \textbf{a}. The lower panels in each subfigure show the MAE between slices of the true and predicted two-time maps as a function of time. Black dotted lines denote the maximum time seen during training. \textbf{d.} Extracted values of the correlation time $\tau$ and Kohlrausch–Williams–Watts exponent $\gamma$ obtained by fitting one-time slices of the true (blue) and predicted (yellow) two-time correlations in Figure \ref{fig:fig2}\textbf{k} with the function  $y=\exp{\left[-\left(\Delta t/\tau\right)^{\gamma}\right]}$, where $\Delta t = |t_1 - t_2|$. \textbf{e-f.} Corresponding values of $\tau$ and $\gamma$ obtained by fitting one-time slices of the true and predicted two-time correlations in \textbf{b} and \textbf{c}, respectively.}}
    \label{fig:figS4}
\end{figure}

\begin{figure}[t]
    \centering
    \includegraphics[width=\textwidth]{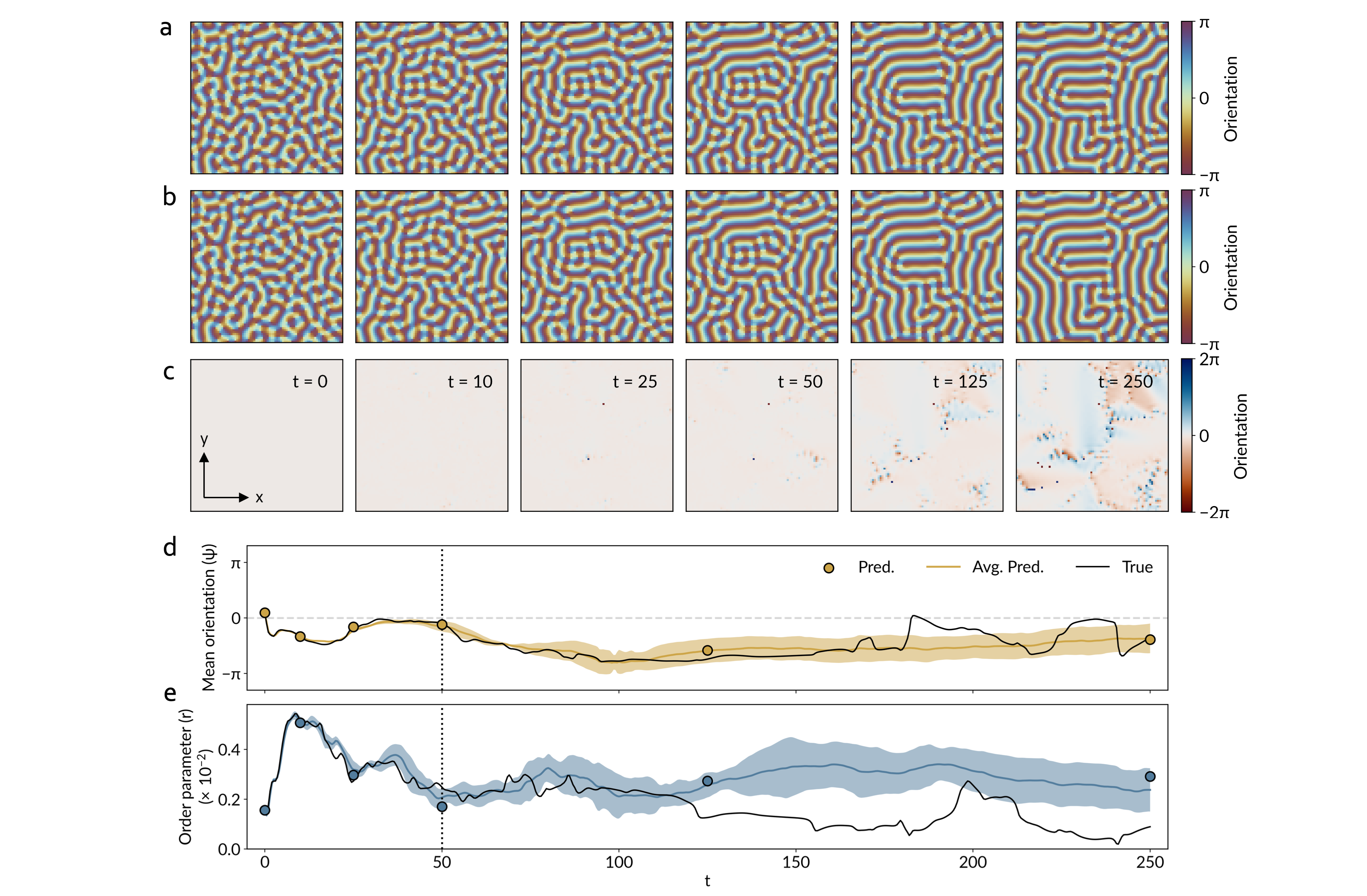}
    \caption{\hlc{\textbf{Real-space analysis for a less challenging initial configuration.} \textbf{a-c.} From top to bottom, the time evolution of the true state, predicted state, and difference map of the system at selected time points for one representative model and initial condition. The final time point ($t=250$) corresponds to five times the maximum time seen during training. \textbf{d-e.} The time evolution of the mean orientation, $\psi$ (\textbf{d}), and order parameter, r (\textbf{e}), of the system in \textbf{a} and \textbf{b}. Solid colored lines correspond to the model ensemble average, and shading represents one standard deviation. Scattered points correspond to the 6 panels in \textbf{b}.}}
    \label{fig:figS5}
\end{figure}

\begin{figure}[t]
    \centering
    \includegraphics[width=\textwidth]{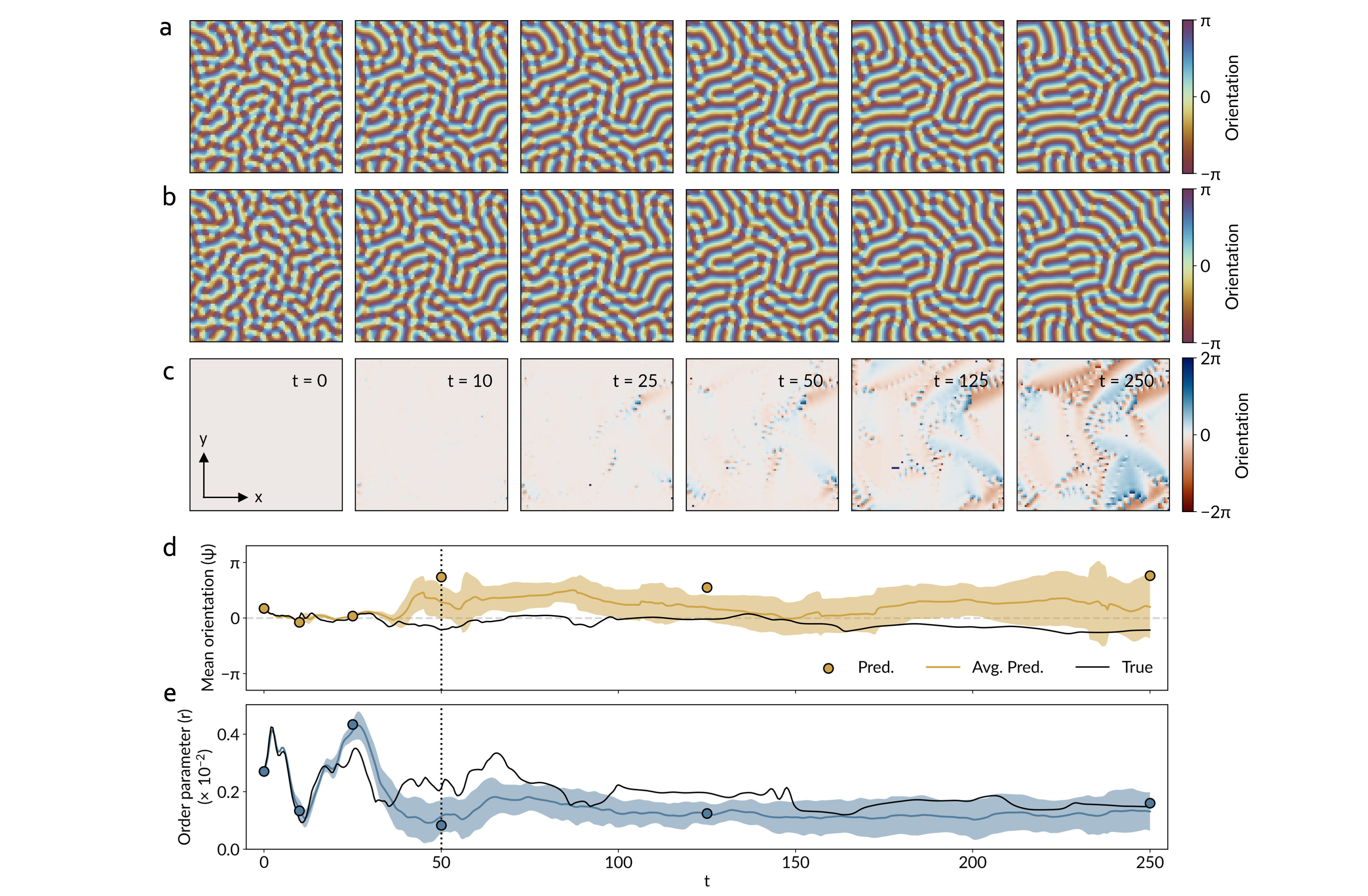}
    \caption{\hlc{\textbf{Real-space analysis for a more challenging initial configuration.} \textbf{a-c.} From top to bottom, the time evolution of the true state, predicted state, and difference map of the system at selected time points for one representative model and initial condition. The final time point ($t=250$) corresponds to five times the maximum time seen during training. \textbf{d-e.} The time evolution of the mean orientation, $\psi$ (\textbf{d}), and order parameter, r (\textbf{e}), of the system in \textbf{a} and \textbf{b}. Solid colored lines correspond to the model ensemble average, and shading represents one standard deviation. Scattered points correspond to the 6 panels in \textbf{b}.}}
    \label{fig:figS6}
\end{figure}

\begin{figure}[t]
    \centering
    \includegraphics[width=\textwidth]{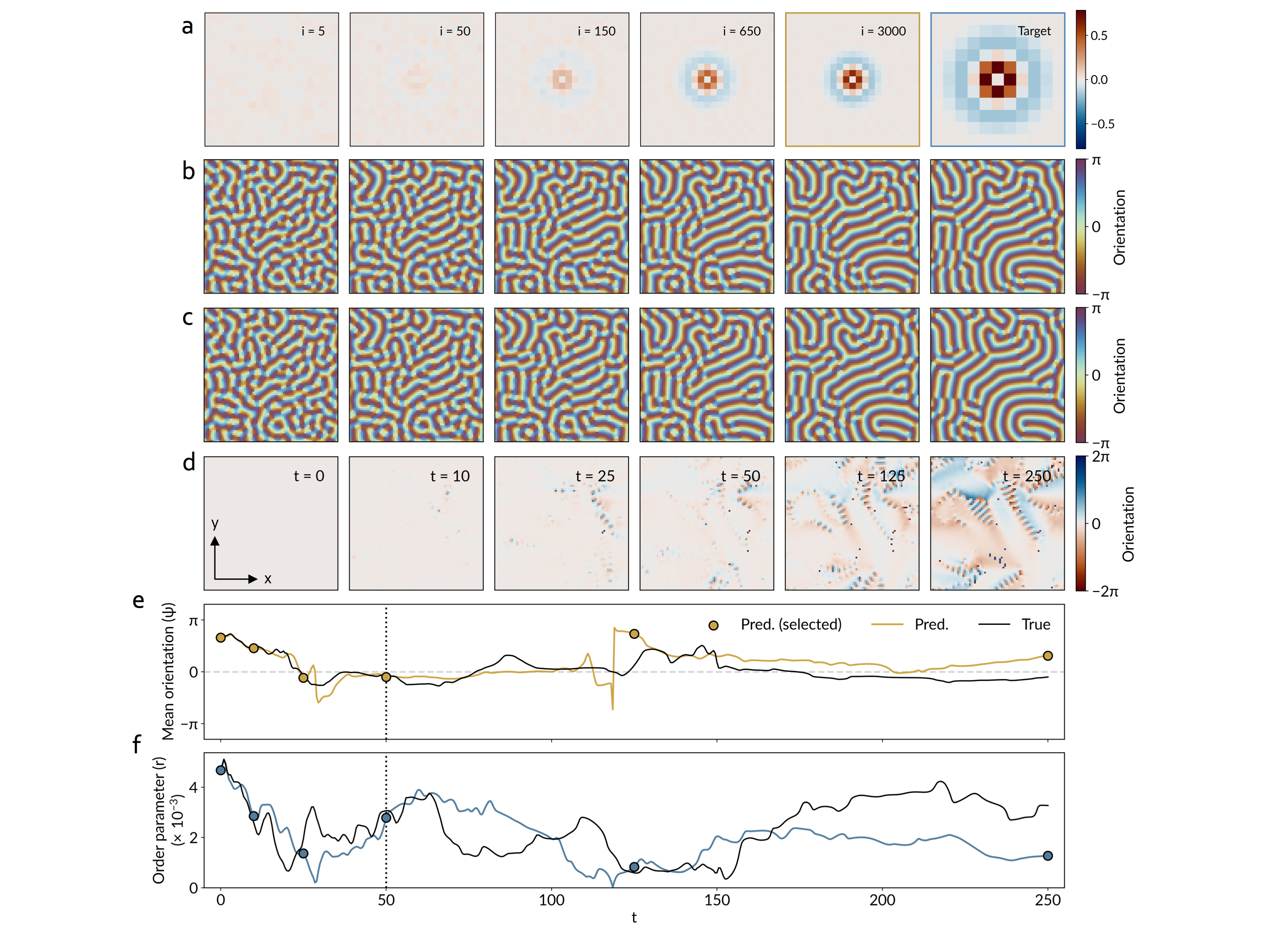}
    \caption{\hlc{\textbf{Overparameterizing the coupling kernel.} \textbf{a.} Evolution of the learned spatial coupling kernel over the course of training, depicted after 5, 50, 150, 650, and 3000 iterations. Panels outlined in yellow and blue represent the final predicted and ground truth kernels, respectively. \textbf{b-d.} From top to bottom, the time evolution of the true state, predicted state, and difference map of the system at selected time points for one initial condition. The final time point ($t=250$) corresponds to five times the maximum time seen during training. \textbf{e-f.} The time evolution of the mean orientation, $\psi$ (\textbf{d}), and order parameter, r (\textbf{e}), of the system in \textbf{b} and \textbf{c}. Scattered points correspond to the 6 panels in \textbf{c}.}}
    \label{fig:figS7}
\end{figure}

\begin{figure}[t]
    \centering
    \includegraphics[width=\textwidth]{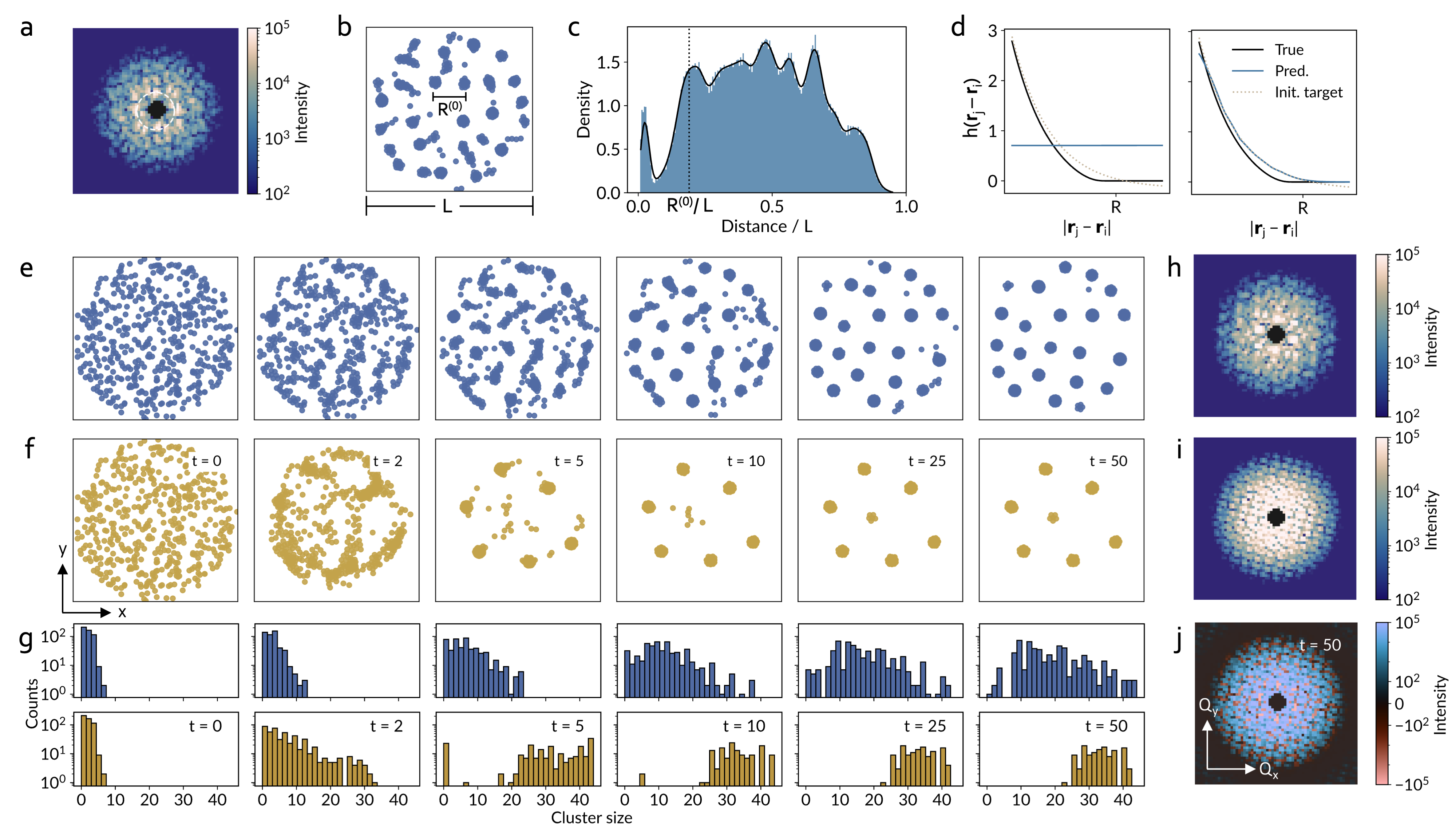}
    \caption{\textbf{Pretraining the model for interacting particles.} \textbf{a.} Representative diffraction pattern at $t=10$, the maximum time point seen during training. The white dashed circle highlights the emergent bright ring of diffraction spots close to the central peak. \textbf{b.} The real-space image corresponding to \textbf{a}. The relevant length scales $L$ and $R^{(0)}$ are indicated. \textbf{c.} Histogram of all pairwise particle distances in \textbf{b}. The vertical dashed line is drawn at the length scale associated with the white dashed circle in \textbf{a}, denoted $R^{(0)}$, which is found to approximately coincide with the average spacing between neighboring clusters in \textbf{b}. \textbf{d.} The predicted cutoff function (blue line) before (left) and after (right) pretraining to the initial target function (dotted line). The true cutoff function is indicated in black. \textbf{\hlc{e-f}.} The time evolution of the \hlc{true (\textbf{e}) and} predicted \hlc{(\textbf{f})} states of the system at selected time points, using only the pretrained cutoff function in \textbf{d} \hlc{for the predicted results}. \textbf{g.} The corresponding time evolution of the true (top) and predicted (bottom) cluster size distributions. \textbf{h-i.} The \hlc{true (\textbf{h}) and} predicted (\hlc{\textbf{i}}) diffraction pattern at $t = 50$. \textbf{j.} The difference map between true and predicted diffraction patterns at $t = 50$.}
    \label{fig:figS8}
\end{figure}

\begin{figure}[t]
    \centering
    \includegraphics[width=\textwidth]{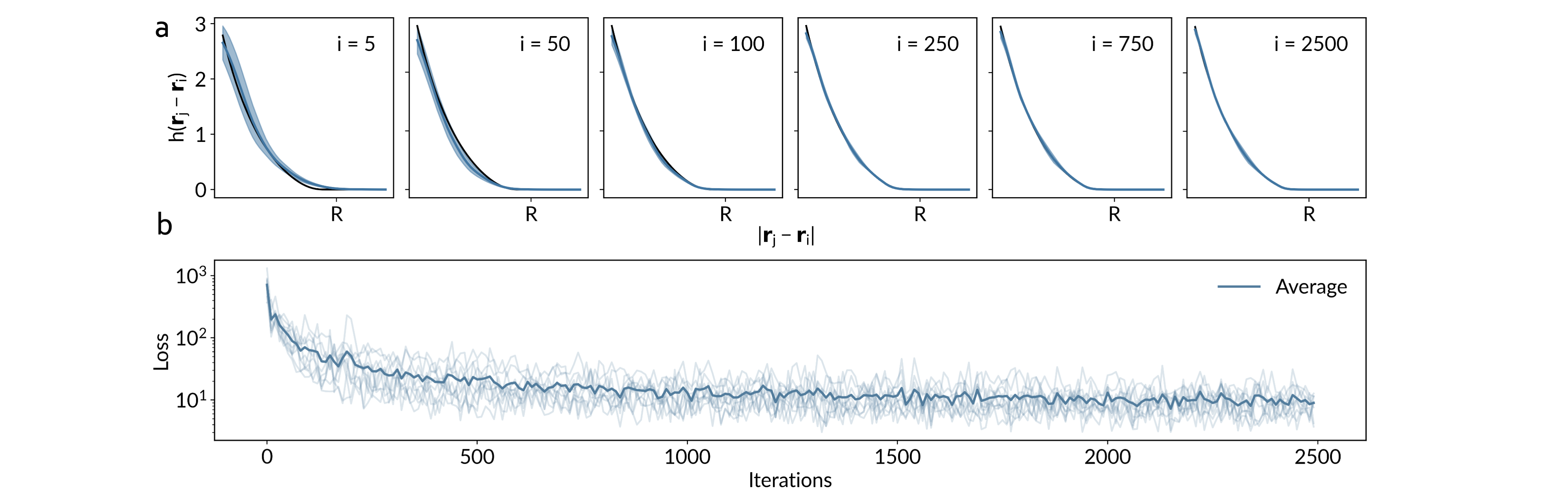}
    \caption{\hlc{\textbf{Training history for a 10-model ensemble.} \textbf{a.} Evolution of the average learned cutoff function during training (after pre-training) Solid blue line shows the average of the 10-model ensemble while the light blue shading represents one standard deviation. The solid black line is the ground truth cutoff function. \textbf{b.} The training loss, computed as the MAE between the true and predicted diffraction patterns, as a function of training time for each of the 10 models (light blue lines) and their average (bold blue line).}}
    \label{fig:figS9}
\end{figure}

\begin{figure}[t]
    \centering
    \includegraphics[width=\textwidth]{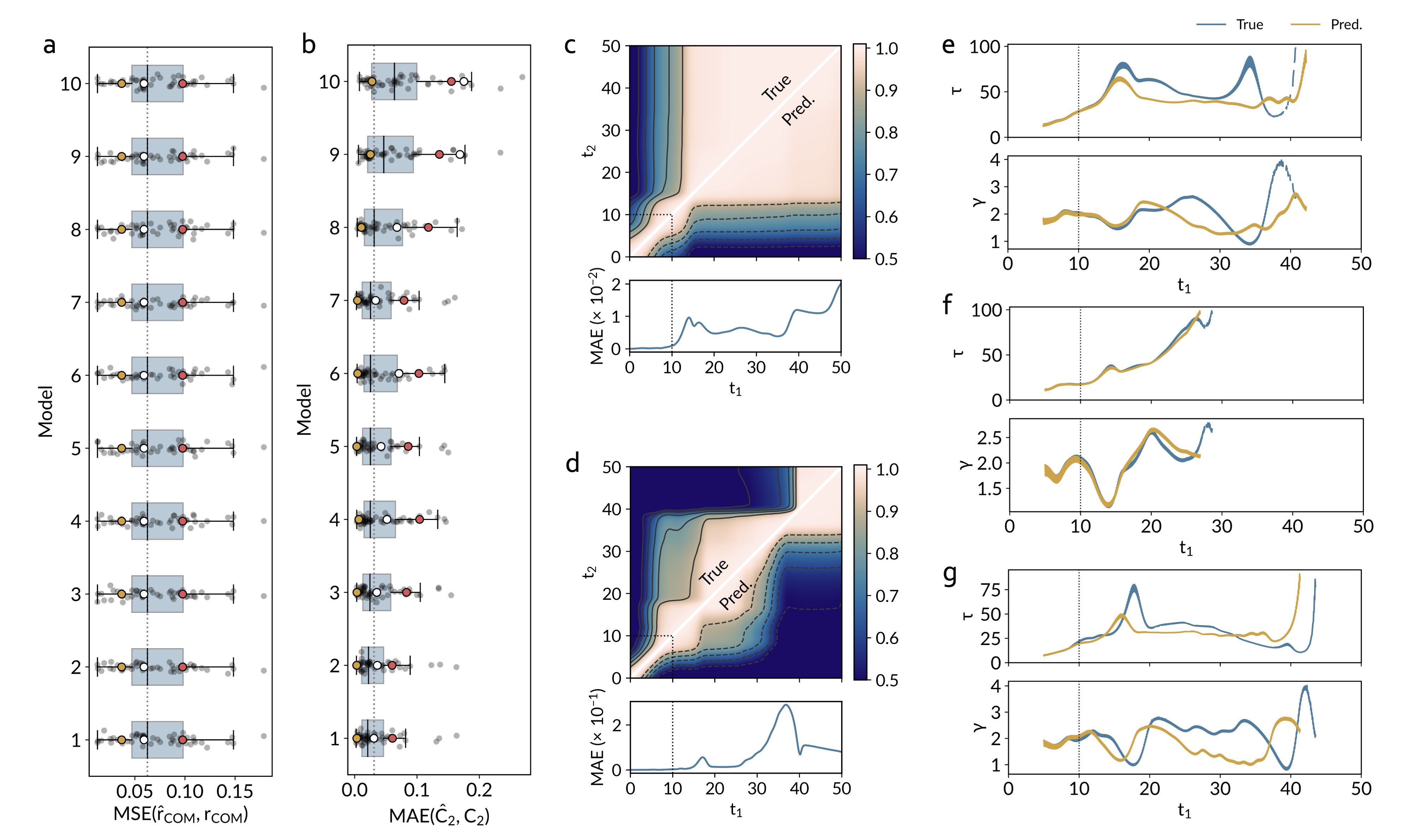}
    \caption{\hlc{\textbf{Two-time correlation function analysis.} \textbf{a.} Boxplots of the MAE between true and predicted centers of mass for each of the 10 models in the ensemble for 50 different initial configurations of the system. Light gray points correspond to each individual data point (initial configuration). Three highlighted data points in each boxplot correspond to the three examples visualized in Figure \ref{fig:fig3} (white), Figure \ref{fig:figS11}\textbf{a}-\textbf{f} (yellow), and Figure \ref{fig:figS11}\textbf{g}-\textbf{l} (red). \textbf{b.} Boxplots of the MAE between true and predicted two-time correlation functions for each of the 10 models in the ensemble for 50 different initial configurations of the system. Light gray points correspond to each individual data point (initial configuration). Three highlighted data points in each boxplot correspond to the same three examples referenced in \textbf{a}. \textbf{c-d.} The two-time intensity-intensity correlation functions calculated from the diffraction patterns of the true (upper triangle) and average predicted (lower triangle) systems for the initial configurations given by the yellow (\textbf{c}) and red (\textbf{d}) points in \textbf{a}. The lower panels in each subfigure show the MAE between slices of the true and predicted two-time maps as a function of time. Black dotted lines denote the maximum time seen during training. \textbf{e.} Extracted values of the correlation time $\tau$ and Kohlrausch–Williams–Watts exponent $\gamma$ obtained by fitting one-time slices of the true (blue) and predicted (yellow) two-time correlations in Figure \ref{fig:fig3}\textbf{c} with the function  $y=\exp{\left[-\left(\Delta t/\tau\right)^{\gamma}\right]}$, where $\Delta t = |t_1 - t_2|$. \textbf{d-g.} Corresponding values of $\tau$ and $\gamma$ obtained by fitting one-time slices of the true and predicted two-time correlations in \textbf{c} and \textbf{d}, respectively.}}
    \label{fig:figS10}
\end{figure}

\begin{figure}[t]
    \centering
    \includegraphics[width=\textwidth]{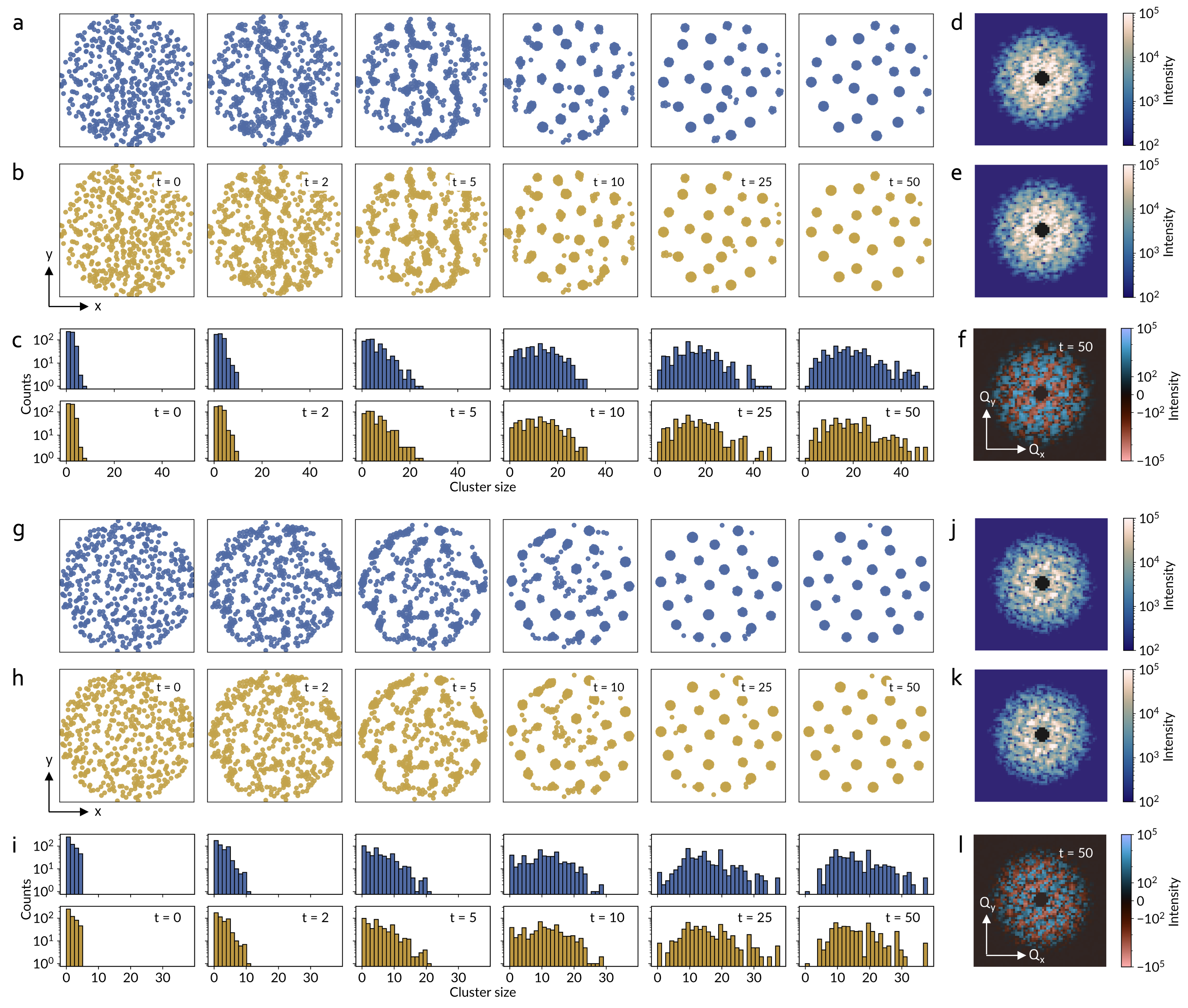}
    \caption{\hlc{\textbf{Real-space analysis for additional initial configurations.} \textbf{a-b.} The time evolution of the true (\textbf{a}) and predicted (\textbf{b}) states of the system at selected time points for one representative model at the initial condition denoted by the yellow point in Figure 
    \ref{fig:figS10}\textbf{a}. The final time point ($t = 50$) corresponds to five times the maximum time seen during training. \textbf{c.} The corresponding time evolution of the true (top) and predicted (bottom) cluster size distributions. \textbf{d-f.} From top to bottom, the true (\textbf{d}) and predicted (\textbf{e}) diffraction patterns and their difference map (\textbf{f}) corresponding to the system state at $t = 50$. \textbf{g-h.} The time evolution of the true (\textbf{g}) and predicted (\textbf{h}) states of the system at selected time points for one representative model at the initial condition denoted by the red point in Figure 
    \ref{fig:figS10}\textbf{a}. The final time point ($t = 50$) corresponds to five times the maximum time seen during training. \textbf{i.} The corresponding time evolution of the true (top) and predicted (bottom) cluster size distributions. \textbf{j-l.} From top to bottom, the true (\textbf{j}) and predicted (\textbf{k}) diffraction patterns and their difference map (\textbf{l}) corresponding to the system state at $t = 50$.}}
    \label{fig:figS11}
\end{figure}

\begin{figure}[t]
    \centering
    \includegraphics[width=\textwidth]{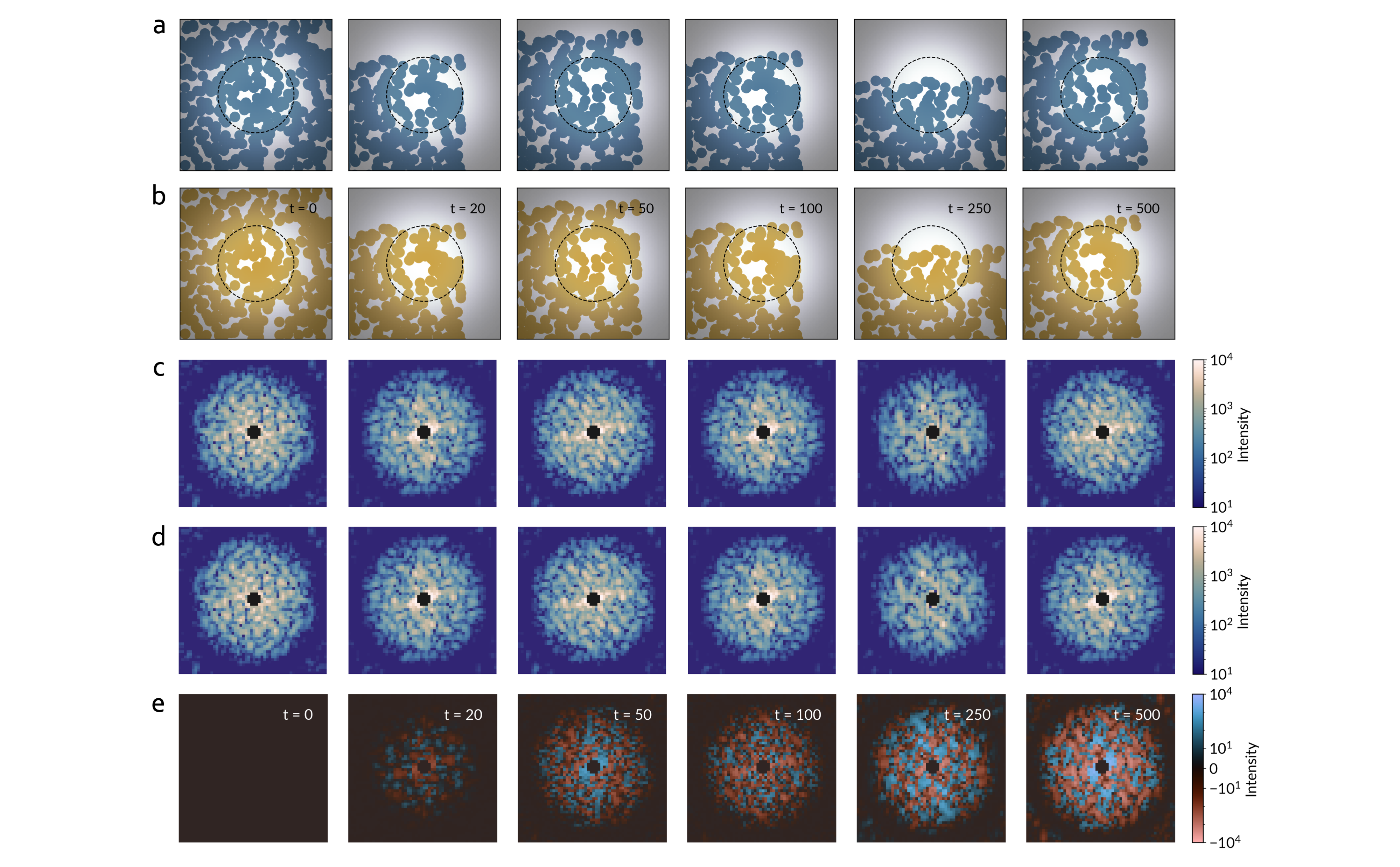}
    \caption{\textbf{Extended results for a fluctuating source.} \textbf{a.} Representative time series of the evolving field-of-view as a result of true source fluctuation. The \hlc{black} dashed circle denotes one standard deviation from the maximum of the Gaussian illumination function. \textbf{b.} The time-evolving field-of-view as a result of predicted source fluctuation. \textbf{c-e.} The true (\textbf{c}) and predicted (\textbf{d}) diffraction patterns and their difference map (\textbf{e}) corresponding to the time series in \textbf{a} and \textbf{b}.}
    \label{fig:figS12}
\end{figure}

\begin{figure}[t]
    \centering
    \includegraphics[width=\textwidth]{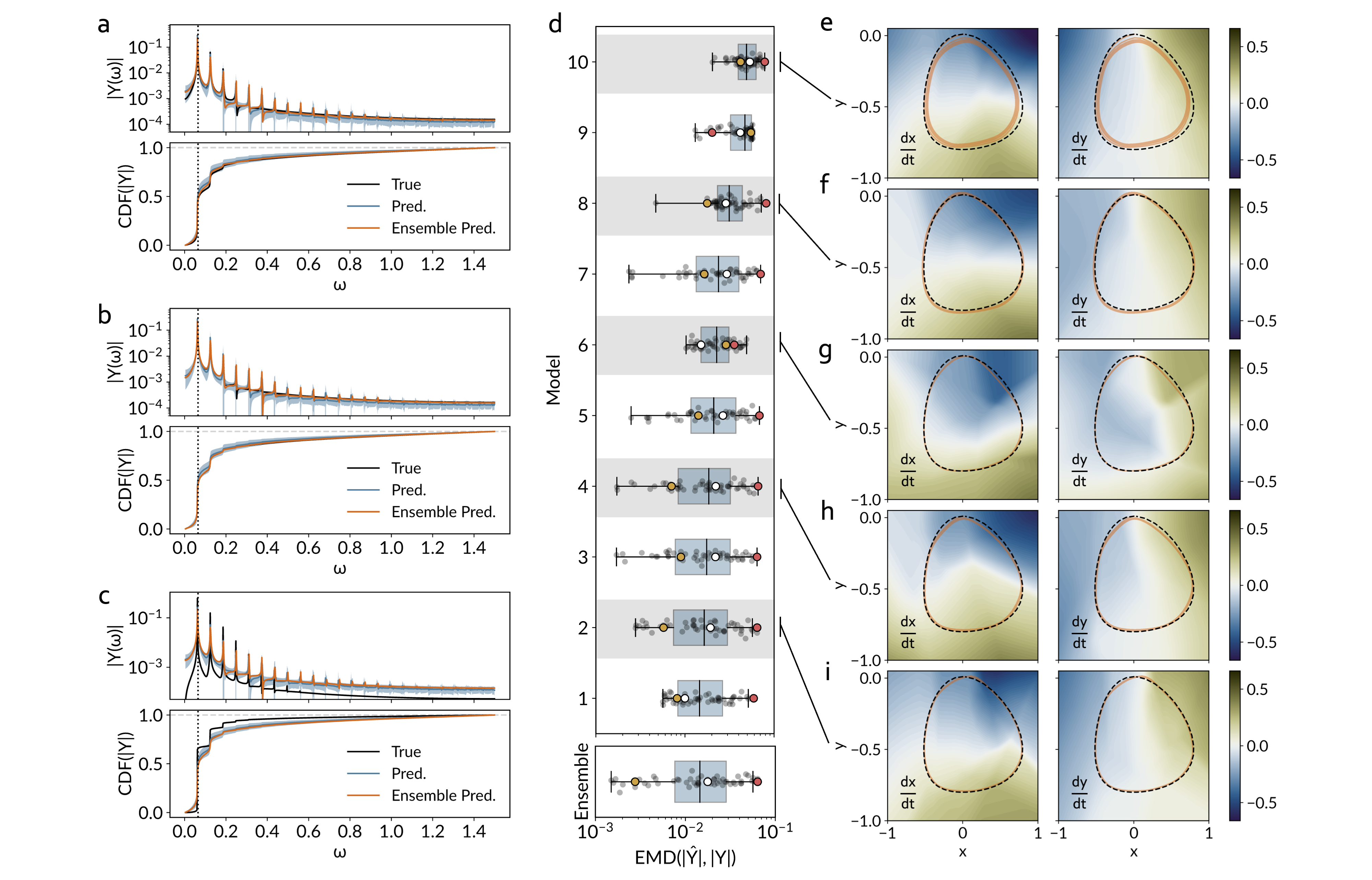}
    \caption{\hlc{\textbf{Extended analysis of learned governing equations.} \textbf{a-c.} Fourier spectra (top panel) and corresponding cumulative distribution functions (CDFs) of the true, predicted, and ensemble-model predicted trajectories for representative simulations with average (\textbf{a}), good (\textbf{b}), and poor (\textbf{c}) performance across the 10 models, as given by the white, yellow, and red scattered points in \textbf{d}, respectively. Blue shading represents one standard deviation across the 10 models in the ensemble. The black vertical dotted lines indicate the analytical frequency of $\sqrt{\alpha \gamma}/2 \pi \approx 0.065$ as described in the main text. \textbf{d.} Boxplots of the Earth Mover's Distance (EMD) errors between true and predicted Fourier spectra for each of the 10 models in the ensemble as well as the ensemble model (bottom panel) for 50 different initial configurations of the system. Light gray points correspond to each individual data point (initial configuration). Three highlighted data points in each boxplot correspond to the three examples visualized in Figure \ref{fig:fig4} (white), Figure \ref{fig:figS14}\textbf{a} (yellow), and Figure \ref{fig:figS14}\textbf{b} (red). \textbf{e-i.} Learned governing equations of the object coordinates for selected models, indicated by the gray rectangles in \textbf{d}, plotted over the domain of interest, with the predicted object trajectory of one representative simulation plotted in orange. The true object trajectory is replotted as a black dashed line for reference.}}
    \label{fig:figS13}
\end{figure}

\begin{figure}[t]
    \centering
    \includegraphics[width=\textwidth]{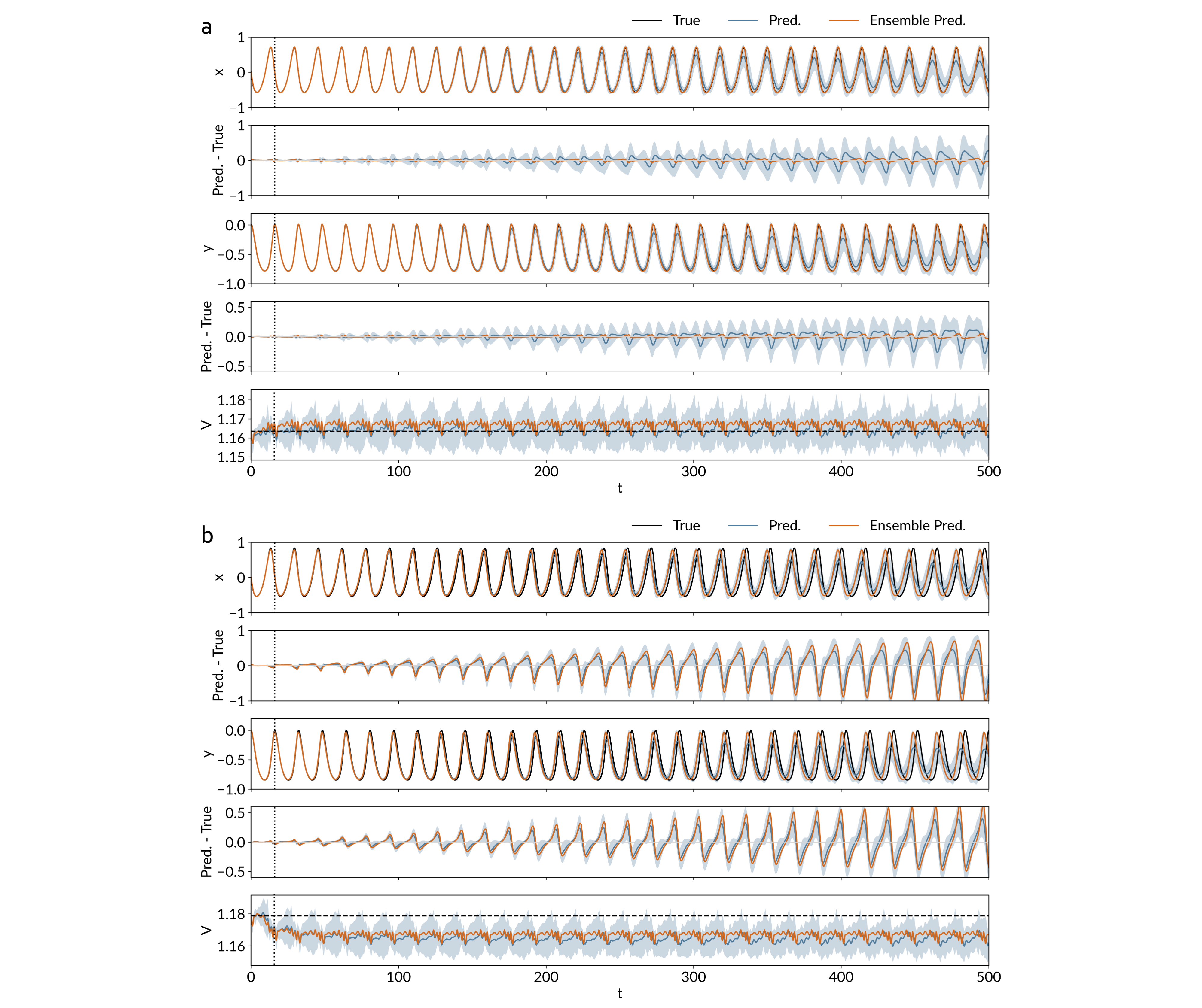}
    \caption{\hlc{\textbf{Real-space analysis for additional initial configurations.} \textbf{a-b.} From top to bottom, the true and predicted object coordinates in the $x$ direction (row 1) and their difference (row 2); the true and predicted object coordinates in the $y$ direction (row 3) and their difference (row 4); and the value of the conserved quantity $V$ (row 5), for the initial condition denoted by the yellow (\textbf{a}) and red (\textbf{b}) scattered points in Figure \ref{fig:figS13}\textbf{d}. The dotted vertical line indicates the maximum time seen during training, $t=16$, while the dashed black lines in rows 5 denote the true values of $V$ in each example. Blue shading denotes one standard deviation across the results of the 10-model ensemble.}}
    \label{fig:figS14}
\end{figure}

\begin{figure}[t]
    \centering
    \includegraphics[width=\textwidth]{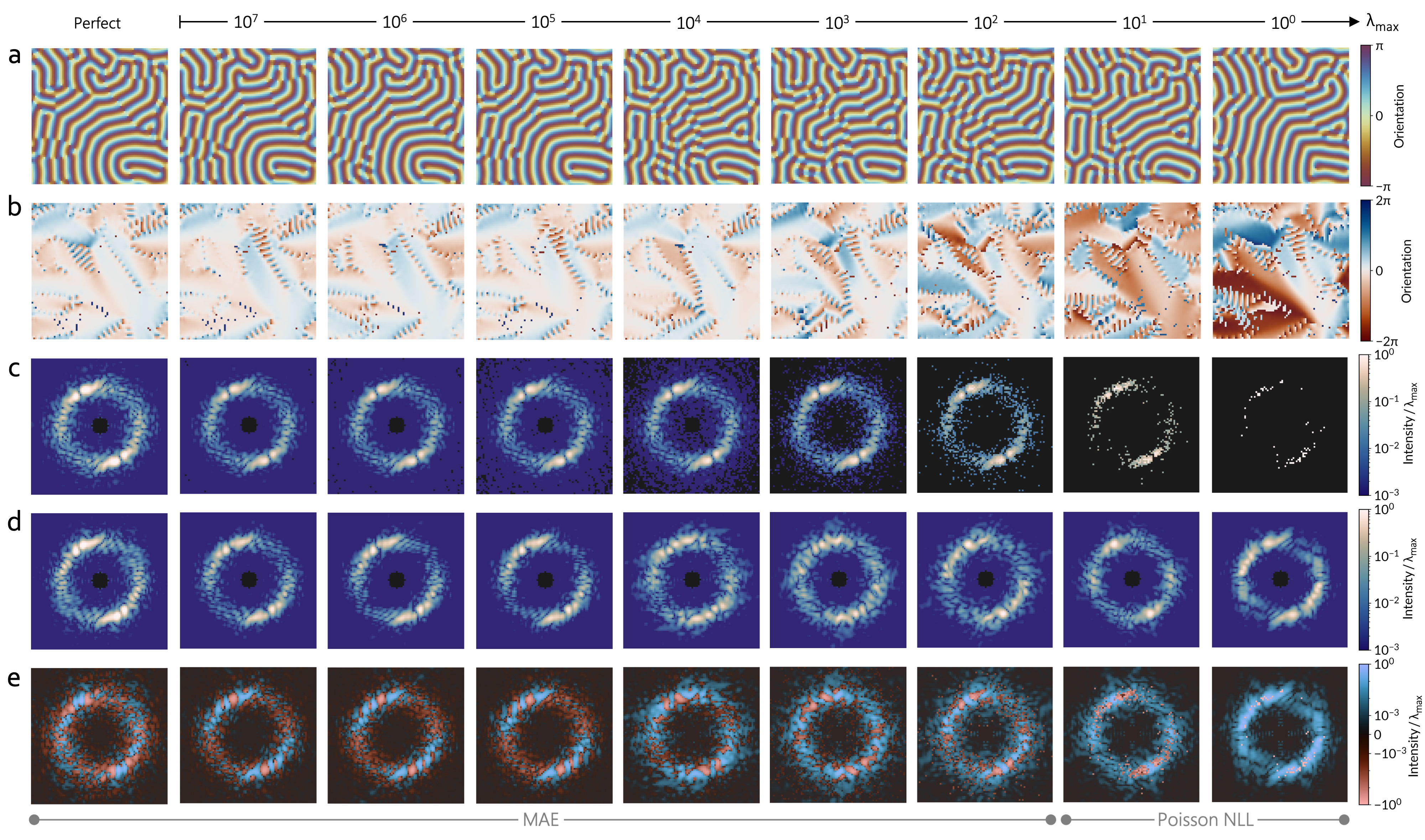}
    \caption{\textbf{Extended effects of measurement noise.} \textbf{a.} Predicted real-space images of the system at $t=250$ as a function of noise, parameterized by $\lambda_{\text{max}}$. The leftmost column corresponds to the perfect case without noise. \textbf{b.} Difference map between the true and predicted real-space images at $t=250$. \textbf{c.} True diffraction pattern calculated at $t=250$ under various levels of simulated noise. \textbf{d.} Predicted diffraction pattern at $t=250$. \textbf{e.} Difference map between \textbf{c} and \textbf{d}.}
    \label{fig:figS15}
\end{figure}

\begin{figure}[t]
    \centering
    \includegraphics[width=\textwidth]{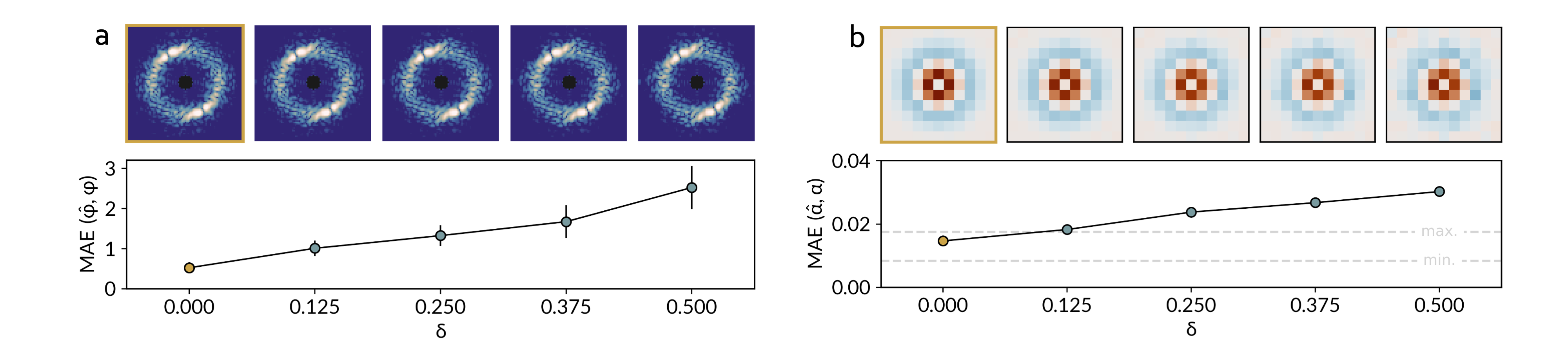}
    \caption{\hlc{\textbf{Effects of form factor perturbation} \textbf{a.} Mean absolute error (MAE) between the true and predicted real-space coordinates at the final extrapolated frame, $t=250$, as a function of the degree of form factor perturbation, $\delta$, which modifies the original form factor $f^{(1)}(\varphi_j)$ according to $f^{(1)}(\varphi_j) = (1 - \delta) \cos{(\varphi_j)} + \delta (\sin^3{(\varphi_j)} + \cos^3{(\varphi_j)})$. The results are an average over 50 different initial conditions, with error bars denoting one standard deviation. Representative diffraction patterns are displayed above each condition. The original condition used for this case study is distinguished by a yellow marker and border around the representative diffraction image. \textbf{b.} MAE between the true and learned coupling kernels as a function of $\delta$. Images of the learned kernels are displayed above each condition. The original condition used for this case study is distinguished by a yellow marker and border around the representative diffraction image. Gray dashed lines denote the minimum and maximum MAEs found for the 10-model ensemble under perfect conditions.}}
    \label{fig:figS16}
\end{figure}

\begin{figure}[t]
    \centering
    \includegraphics[width=\textwidth]{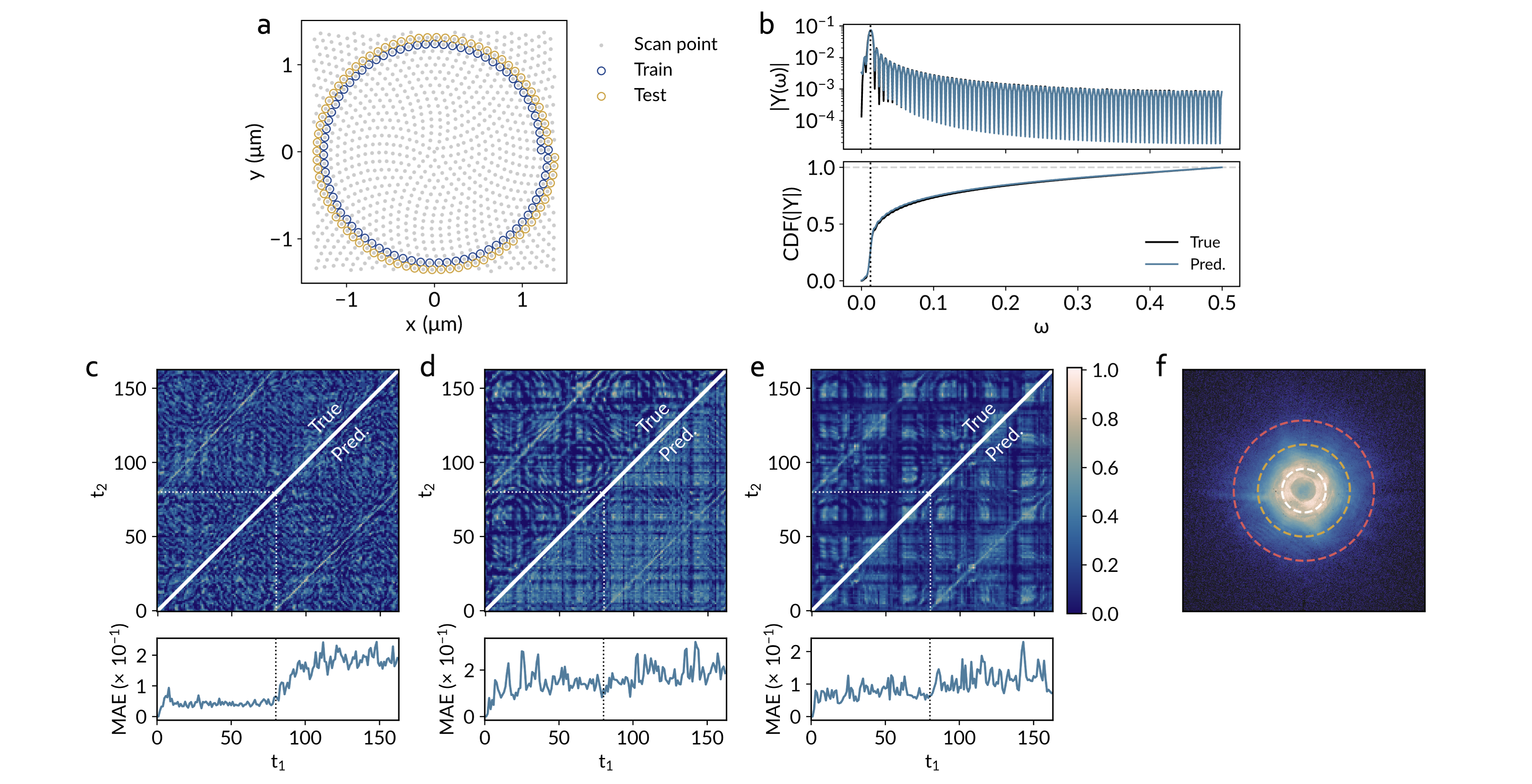}
    \caption{\hlc{\textbf{Extended analysis for experimental proof-of-concept.} \textbf{a.} Plot of the scan points comprising the full ptychography experiment. The points selected for training and testing (extrapolation) are outlined in blue and yellow, respectively. \textbf{b.} Fourier spectra (top panel) and corresponding cumulative distribution functions (CDFs) of the true and predicted trajectories. The black vertical dotted lines indicate the true frequency of $1/80 = 0.0125$ as described in the main text. \textbf{c-e.} The two-time intensity-intensity correlation functions calculated from the diffraction patterns of the true (upper triangle) and predicted (lower triangle) systems. The correlation functions were calculated by taking an azimuthal average over each diffraction pattern at the equivalent $Q$ indicated by the white (\textbf{c}), yellow (\textbf{d}), and red (\textbf{e}) dashed lines in \textbf{f}. \textbf{f.} Representative reciprocal-space diffraction pattern with $Q$ values for the two-time correlation analysis indicated as described in \textbf{d-e}.}}
    \label{fig:figS17}
\end{figure}

\begin{figure}[t]
    \centering
    \includegraphics[width=\textwidth]{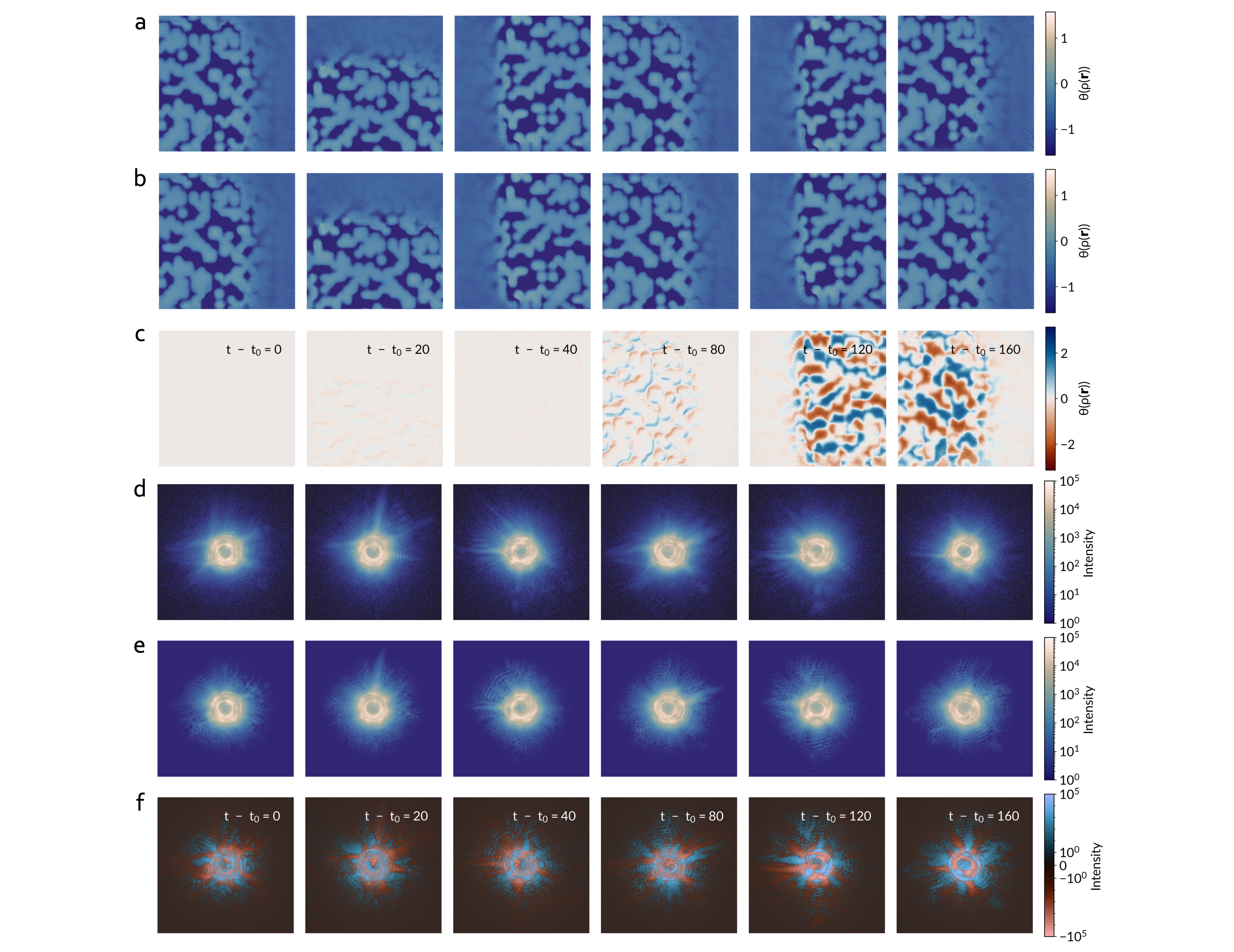}
    \caption{\textbf{Extended results for experimental proof-of-concept.} \textbf{a-c.} The time series of the real-space phase along the \textbf{a} true and \textbf{b} predicted probe trajectory and their difference map (\textbf{c}). \textbf{d-f.} The true (\textbf{c}) and predicted (\textbf{d}) diffraction patterns and their difference map (\textbf{e}) corresponding to the time series in \textbf{a} and \textbf{b}.}
    \label{fig:figS18}
\end{figure}

\end{document}